\documentclass[a4paper, authoryear, 11pt]{article}

\usepackage[english]{babel}
\usepackage{pifont}
\usepackage{color}
\usepackage{babel}
\usepackage{fontenc}
\usepackage{graphicx}
\usepackage{natbib}
\usepackage[colorlinks=true,citecolor=blue]{hyperref}
\usepackage[usenames,dvipsnames]{pstricks}
\usepackage{amsmath,amsthm, amsfonts, amssymb, mathrsfs,hyperref}
\usepackage{epsfig}
\usepackage{amsthm}
\usepackage{multicol}
\usepackage{float}
\usepackage{caption}
\usepackage{subcaption}

\numberwithin{equation}{section}

\providecommand{\keywords}[1]
{
  \small	
  \textbf{\textit{Keywords---}} #1
}

\title{The balance between contamination and predation determine species existence in prey-predator dynamics with contaminated and uncontaminated prey  }
\author{ Amit Samadder, Arnab Chattopadhyay, Sabyasachi Bhattacharya \footnote{Agricultural and Ecological research Unit, Indian Statistical Institute, 203, B.T Road, Kolkata- 700108, India. E-mail: math.amitsamadder18@gmail.com, arnabchatterjee891@gmail.com, sabyasachi@isical.ac.in}
\footnote{Corresponding Author, Contact No. - (+91)-9433897120, Fax:(91)(33) 2577-3049}}

\begin{document}
\maketitle
\begin{abstract}
In freshwater ecosystems, aquatic insects that ontogenetically shift their habitat from aquatic to terrestrial play vital roles as prey subsidies that move nutrients and energy from aquatic to terrestrial food webs. As a result, these subsidies negatively affect alternative terrestrial prey by enhancing predator density. However, these aquatic insects can also transport contamination to the terrestrial community that is primarily produced in aquatic ecosystems. Which can reduce insectivore abundance and biomass, lower insectivore reproductive success, and increase predation pressure on alternative prey with consequences for aquatic and terrestrial food webs. Motivated by this, here we consider a prey-predator model where the predator consumes contaminated and uncontaminated prey together. We find that, at a high level of contamination, the vulnerability of contaminated prey and predator is determined by predation preference. More specifically, a very low predation preference towards contaminated prey ensures predator persistence, whereas a low, high preference excludes the predator from the system. Interestingly, either contaminated prey or the predator exist at intermediate predation preference due to bi-stability.
Furthermore, when there is no contamination in one of the prey, the other prey can not coexist due to apparent competition for a specific range of predation preferences. However, when sufficient contamination exists in one prey, alternative uncontaminated prey coexists. With this, contamination also stabilizes and destabilizes the three species dynamics. Our result also indicates that if the intensity of the contamination in predator reproduction is low, then contaminated prey is more susceptible to the contamination.                
\end{abstract}
\newpage
\section{Introduction}\label{Introduction}
Across the globe is growing concern about environmental pollution in ecosystem health, as it is one of the direct drivers of ecosystem change at the global scale and a significant contributor to biodiversity loss \citep{nelson2005drivers,hooper2012global}. Pollution can adversely affect a diverse range of organisms by impairing their neurological function, behavior, and life cycle parameters, like survival and reproduction \citep{wu2016review,wolfe1998effects}. Organisms exposed to the polluted environment often take up contaminants through food and water. These contaminants are then passed onto the higher trophic level through trophic interactions and consequently spread toxicity across the the food web. So, quantifying the risk of pollutants requires not only a prior knowledge of their effect on individual organisms but also an understanding of how interacting species respond to chemical stressors.
\newline
Several theoretical studies have examined the impact of environmental contamination on ecosystem stability, persistence\citep{hallam1983effects,zhien1990persistence,baudrot2018effects,huang2015impact, prosnier2015modeling, kooi2008sublethal}. The earliest mechanistic models in this area were the works based on single species dynamics\citep{hallam1983effects,zhien1990persistence}. Further, it was extended for the trophic interactions. For instance, \cite{huang2015impact} predicts that high concentrations of methylmercury lead to the collapse of the entire bi-trophic food chain, while at intermediate toxin concentrations, predators become more vulnerable than the contaminated prey. \cite{prosnier2015modeling} also point out that the consumer population is more susceptible to less extreme copper concentrations than the contaminated prey. Apart from the trophic food chain, the impact of contamination on complex, large food webs is also done and shows that in a more polluted environment, topological properties of the food web are more vital for stability \citep{garay2013more, garay2014food}.   
\newline
Most of these theoretical studies focus on the impact of contamination on interacting species dynamics by considering all the prey of the predator is contaminated. However, in a natural ecosystem, many generalist predators can consume contaminated and uncontaminated prey together. For instance, \cite{palma2005spatial} find that Bonelli eagles acquire most of their burden through the intake of secondary consumers as mercury levels were very low in alternative prey. Further, some consumers tend to forage over larger spatial scales even up to kilometers, naturally they can consume prey from contaminated and uncontaminated sites. Again, prey that disperses from contaminant sites can transport contaminants to a spatially separated prey-predator community\citep {paetzold2011environmental,kraus2019contaminants,bartrons2015taking,schiesari2018metacommunities}. Recent evidence demonstrates that aquatic insects export substantial mercury from the aquatic environment to terrestrial insectivore communities, like songbirds, which frequently consume terrestrial and aquatic insects\citep {cristol2008movement,jackson2011mercury,jackson2021differential}. However, we know very little about how contamination affects these communities.
\newline
Theory and experimental evidence predict that prey subsidies can support a denser predator population, consequently increase predation on the alternative prey population\citep{holt1977predation,polis1997toward,baxter2005tangled,murakami2002indirect, henschel2001allochthonous}. Despite the benefits of aquatic insect emergence as an alternative energy source to the predator, they can also degrade predators through the tropic transfer of aquatically derived environmental contaminants\citep{kraus2019contaminants,schiesari2018metacommunities,manning2021conservation}, such as methylmercury, that primarily produced in aquatic environments \citep{selin2009global,hall2008wetlands}. Indeed, in the face of current environmental pollution, predators are predicted to be more susceptible than the contaminated prey\citep{huang2015impact, prosnier2015modeling}. However, it is difficult to predict the victim of contamination when a predator accumulate contaminates through the prey only and also have an alternative uncontaminated energy source,  because, while contamination negatively affects both the predator and contaminated prey, the flow of contamination to the predator also depends on the predator's diet composition\citep{jackson2021differential,ortega2019relationship,palma2005spatial}. So the predators more sensitive to the contamination may not suffer from considerable extinction risk if they rarely prefer to consume contaminated prey. Conversely, predators less sensitive to contamination can carry high extinction risk if they frequently consume contaminated prey. Furthermore, a high predation pressure can also have a determental effect for the contaminated prey and predator in highly contaminated environment \citep{beketov2006influence}.  
\newline
Apart from the predator, contamination may also regulate indirect interaction between preys. In general, food web modules such as multi-prey with a common predator, prey is known to compete with each other indirectly, also known as apparent competition\citep{polis1997toward}. The idea of apparent competition is two prey species negatively affect each other by increasing the equilibrium density of the shared predator, and in an extreme case, one prey excludes another alternative prey from the community \citep{holt2017apparent}. Experimental evidence also supports this phenomenon by showing that aquatic insects subsidiary suppress terrestrial herbivore insects by increasing predator density \citep{baxter2005tangled,murakami2002indirect,henschel2001allochthonous}. In apparent competition between terrestrial and aquatic emergent insects, contamination in aquatic prey can reduce top-down pressure on the uncontaminated terrestrial prey by decreasing the predator density, thereby can initiate its persistence. Conversely, predators that prefer quality over quantity can also increase predation pressure on uncontaminated prey by shifting their diet to more favorable prey \citep{marcarelli2011quantity}, thus can enhance its extinction risk. \cite{shan2019direct} find that contamination can alter the outcome of the direct competition by suppressing competitively superior species. Nevertheless, how contamination affects indirect competition between prey is yet to be known.          
\newline
In light of the above-known facts, here we formulate a prey-predator model where the predator consumes contaminated and uncontaminated prey together. We mainly attempt to answer the interrelated questions: (1) Does contamination become a precursor of extinction for the predator or the contaminated prey? (2) How does contamination affect the indirect competition between prey? (3) What is the role of predation preference in rescuing or imperiling each species in contaminated environments? We use bifurcation analysis to investigate this system's possible asymptotic states and hypothesize that direct and indirect species interaction have a profound effect in driving the dynamics of contaminated systems.
\newline
The rest of the paper is organized as follows. In section\ref{Methods}, we develop a toxin-dependent prey-predator model with uncontaminated prey. In this section, we also derive dimensionless form by suitable parameter substitutions. In section\ref{Results}, we investigate the possible asymptotic states of our system with the help of bifurcation
diagrams. We also check the robustness of the result for other parameter sets and alternative assumptions. Finlay, in section\ref{Discussion}, we complete the paper with a brief discussion.

\section{Methods}\label{Methods}
\subsection{Modelling framework}
 Our proposed model consists of three populations: $X$, $Y$, and $Z$. We assume that $X$ is the contaminated prey density, $Y$ is the uncontaminated prey density, and $Z$ is the predator population density that consumes both prey with some preferences. We assume contamination pollutes water-based habitat, which enters into the prey $X$ by direct absorption from water. When it comes to terrestrial habitat, terrestrial predators get contaminated by consuming that contaminated prey. We also assume that contaminated prey is the only source of contamination for the predator. We did not assume any direct competition between prey due to their dietary separation. In absence of the contamination in the prey $X$, the growth profile describes by the following set of differential equations:
 \begin{equation}
     \frac{dX}{dt}=r_{1}X(1-X/K_{1})-\frac{aqXZ}{\zeta+qX+(1-q)Y}
 \end{equation}
 \begin{equation}
     \frac{dY}{dt}=r_{2}Y(1-Y/K_{2})-\frac{a(1-q)YZ}{\zeta+qX+(1-q)Y}
 \end{equation}
 \begin{equation}
     \frac{dZ}{dt}=e(\ \frac{aqXZ+a(1-q)YZ}{\zeta+qX+(1-q)Y})\ -mZ
 \end{equation}
 The above model represents an apperent competition model under non-switching framework\citep{faria2009interplay}.
 The parameters $r_{i} (i=1,2)$ denotes the intrinsic growth of the preys, $K_{i} (i=1,2)$ are the carrying capacities of the preys. We assume saturating functional response where $a$ is the attack rate of the predator and $q, \ (1-q), (0\leq q \leq 1)$ are the preferences of the predator for the prey $X$ and $Y$ respectively. $e$ is the conversion efficiency of the predator and $m$ is the natural mortality rate.
 
 \subsubsection{Modelling dynamics of the internal contamination concentration }
 A general assumption in modeling the ecological dynamics of a system and pollution dynamics of the individuals together is that, the dynamics of the pollution in the species body operate much faster timescale than the ecological dynamics. So the internal concentration in an individual's body approaches the steady state before the population dynamics change significantly. Simultaneously, it reduces the model dimensions and facilitates analysis \citep{baudrot2018effects,huang2015impact,prosnier2015modeling,kooi2008sublethal} . Bellow, we derive the internal concentration of the contamination in the individual body. 
 \newline
 A simple and mechanical  bio-dynamic model of internal concentration of contamination can be written by the following form\citep{luoma2005metal,veltman2008cadmium}:
 \begin{equation}
     C_{t}^{'}=(\ I_{W}+I_{F} )\ -k_{e}C_{t}
 \end{equation}
 where $C_{t}$ is the internal concentration of the contamination of an organism's body at time $t$. $I_{W}$ and $I_{F}$ are the unidirectional fluxes of the contamination coming from environment and food respectively. $k_{e}$ is the constant loss rate of the contamination from body of the organism. We can write $I_{W}$, $I_{F}$ as,
 \newline
\begin{equation}
    I_{W}=k_{u}\times C_{c}
\end{equation}
\begin{equation}
    I_{F}=AE\times IR\times C_{F}
\end{equation}
Where $C_{c}$ is the concentration of the contamination  in the environment, $k_{u}$ denotes uptake constant in the body,  AE= assimilation efficiency, IR=egnation rate, $C_{F}$=concentration in food. Combining equation (2.4), (2.5) and (2.6) we get the steady state approximation of internal concentration of contamination by equating $C^{'}_{t}=0$ as a ratio of total uptake divided
elimination rate constant \citep{veltman2008cadmium} ,
\begin{equation}
    C_{ss}=\frac{k_{u}\times C_{c}+AE\times IR\times C_{F}}{k_{e}}
\end{equation}
The internal concentration of the contaminant of the prey is:
\begin{equation}
    C=\frac{k_{u}\times C_{c}}{k_{e}^{prey} }
\end{equation}
The internal contaminant in the predator can be written in terms of the prey contamination concentration(Prey internal concentration) as:
\begin{equation}
    C_{P}=\beta\times \frac{ aqX}{(\ \zeta+qX+(1-q)Y )\ k_{e}^{predator}}\times C
\end{equation}

$\beta$ is the assimilation efficiency of the predator.
 \newline
 \subsubsection{Incorporating contamination in prey predator dynamics}
 Before incorporating contamination in to our three species model we make following assumptions:
 \begin{itemize}
     \item We let the growth of contaminated prey is $\frac{r}{1+kC}$, where $C$ is the pollution in prey  and  $k$ is the intensity of the pollution in the prey growth. We choose such type of effect of contamination in prey growth to avoid the prey extinction due to  contamination only. By doing this we can access other effects that may drive contaminated prey to the extinction. We also assume that contaminated prey have no contamination related mortality(but see  section \ref{Robustness2}).
     \item We consider $Y$ as a terrestrial herbivore which did not consume contamination due to boundary separation between contaminated and uncontaminated sites.
     \item The reproduction of the predator Z directly affected by the consumption of the contaminated prey. We implement the reduction of the predator reproduction by rewriting the conversion efficiency in the form of $max(0,1-bC_{p})=max(0,1-\frac{\beta baqXC}{(\ \zeta+qX+(1-q)Y )\ k_{e}^{predator}})$, $b$ is the intensity of the pollution in the predator reproduction in other words $\frac{1}{b}$ is the internal concentration at which reproduction is zero.
     \item We did not incorporate the effect of the contamination in the predator mortality as songbird is our model predator and studies found that songbird did not exhibit significant difference in the mortality between contaminated and uncontaminated sites\citep{hallinger2011mercury}. 
 \end{itemize}
 
 Keeping above assumptions in mind we can write the contaminated prey predator system as:
 \begin{equation}
     \frac{dX}{dt}=\frac{r_{1}}{1+kC}X(1-X/K_{1})- \frac{aqXZ}{\zeta+qX+(1-q)Y}
 \end{equation}
 \begin{equation}
     \frac{dY}{dt}=r_{2}Y(1-Y/K_{2})-\frac{a(1-q)YZ}{\zeta+qX+(1-q)Y}
 \end{equation}
 \begin{equation}
     \frac{dZ}{dt}=e max(\ 0, 1-\frac{\beta baqXC}{(\ \zeta+qX+(1-q)Y)\ k_{e}^{predator}} )\ (\ \frac{aqXZ+a(1-q)YZ}{\zeta+qY+(1-q)Y} )\ -mZ
 \end{equation}

\begin{table}[H]
\label{table}
 \small
 \caption{Variables and Parameter description}
 \vspace{0.5 cm}
 \label{par_table}       
 \begin{tabular}{lllll}
 \hline\noalign{\smallskip}
 
Parameters                    & Unit             & Description                                                   \\
\noalign{\smallskip}\hline\noalign{\smallskip}
$X$   & $g/m^{3}$  & contaminated prey density \\

$Y$   & $g/m^{3}$  & uncontaminated prey density \\

$Z$   & $g/m^{3}$  & predator density \\
$k_{u}$ &$Lg^{-1}day^{-1}$ & prey absorption rate of the contamination\\
$k_{e}^{prey}$ &$day^{-1}$ & contamination elimination rate of the prey\\

$C$   & $\mu g/g$  & contamination in prey\\
$C_{c}$ &$\mu g/ L$ & contamination in water\\
$k$   & $g/\mu g$  & intensity of contamination in prey growth\\

$r_{1}$   & $day^{-1}$  & contaminated prey intrinsic growth rate \\

$r_{2}$   & $day^{-1}$  & uncontaminated prey intrinsic growth rate \\

$K_{1}$   & $g/m^{3}$  & contaminated prey carrying capacity\\

$K_{2}$   & $g/m^{3}$  & uncontaminated prey carrying capacity\\

$a$   & $day^{-1} $  & attack rate of the predator\\
$\zeta$   & $g/m^{3} $  & half saturation constant \\

$q$   & $-$  & predation preference\\
$e$   & $-$  & conversion efficiency\\
$b$   & $g/\mu g$  & intensity of contamination in predator\\
$m$   & $day^{-1}$  & predator natural mortality rate\\
$\beta$   & $-$  & assimilation efficiency of the predator\\
$k_{e}^{predator}$   & $day^{-1}$  & pollution elimination rate of the predator\\

\noalign{\smallskip}\hline
\end{tabular}
 \end{table}
 \subsubsection{Dimensionless Model}
 To reduce the number of model parameters we dimensionless the above model by considering following substitutions:
 \newline
 $\tau=r_{1}t$,  $x=\frac{X}{K_{1}}$, $y=\frac{y}{K_{1}}$, $z=\frac{Z}{K_{1}}$, $l=\frac{a}{r_{1}}$, $c=kC$, $\eta=\zeta/K_{1}$, $r=\frac{r_{2}}{r_{1}}$ $\lambda=\beta\frac{r_{1}b}{k_{e}^{predator}k}$, $d=\frac{m}{r_{1}}$, $K=\frac{K_{2}}{K_{1}}$
 \newline
 The dimensionless form of our above model is:
 \begin{equation}
     \frac{dx}{d\tau}=\frac{x}{1+c}(1-x)-\frac{lqxz}{\eta+qx+(1-q)y}
 \end{equation}
 \begin{equation}
     \frac{dy}{d\tau}=ry(1-y/K)-\frac{l(1-q)yz}{\eta+qx+(1-q)y}
 \end{equation}
 \begin{equation}
     \frac{dz}{d\tau}=el max(\ 0, 1-\frac{\lambda ql xc}{\eta+qx+(1-q)y} )\ (\ \frac{qxz+(1-q)yz}{\eta+qx+(1-q)y} )\ -dz
 \end{equation}
\subsection{Model calibration and analysis}
Mathematical analysis of positively and boundedness of the solutions carried out in Appendix.A to ensure that our model mechanistically well define.   Standard methods
of linear stability analysis provides the analytical conditions for existence and
stability of the various equilibrium of the system in appendix.B. 
We perform extensive bifurcation analysis to explore possible emerge dynamics of the system (2.13-2.15). Numerical bifurcations are performed in MATLAB using numerical continuation software MATCONT \citep{dhooge2008new}and father verified by tracing time series data trough deSolve package in R. For the lake of sufficient parameter values form existing literatures in the case of our modeled species we assume hypothetical parameter values which are biologically feasible:
$l=1$, $\eta_{1}=0.5$, $\eta_{2}=0.5$, $r=0.5$, $K=1$, $e=0.4$, $\lambda =1.7$, $d=0.1$, parallely we also investigate the dynamics for aditional parameter sets in section \ref{Robustness1}. Throughout the analysis we set parameters such a way that make the species $Y$ as inferior indirect competitor and $X$ as superior indirect competitor in absence of any contamination.

\section{Results}\label{Results}
In this section we used few notations to describe dynamics of the following system. The stable coexistence equilibrium and oscillatory coexistence equilibrium denoted by SC, OC respectively. The contaminated and uncontaminated prey extinction equilibrium denoted by CPE, UPE while the predator extinction equilibrium referred to as PDE.

\subsection{Prey predator dynamics in uncontaminated system}
\begin{figure}[H]
     \centering
     \begin{subfigure}[b]{0.3\textwidth}
         \centering
         \includegraphics[width=\textwidth]{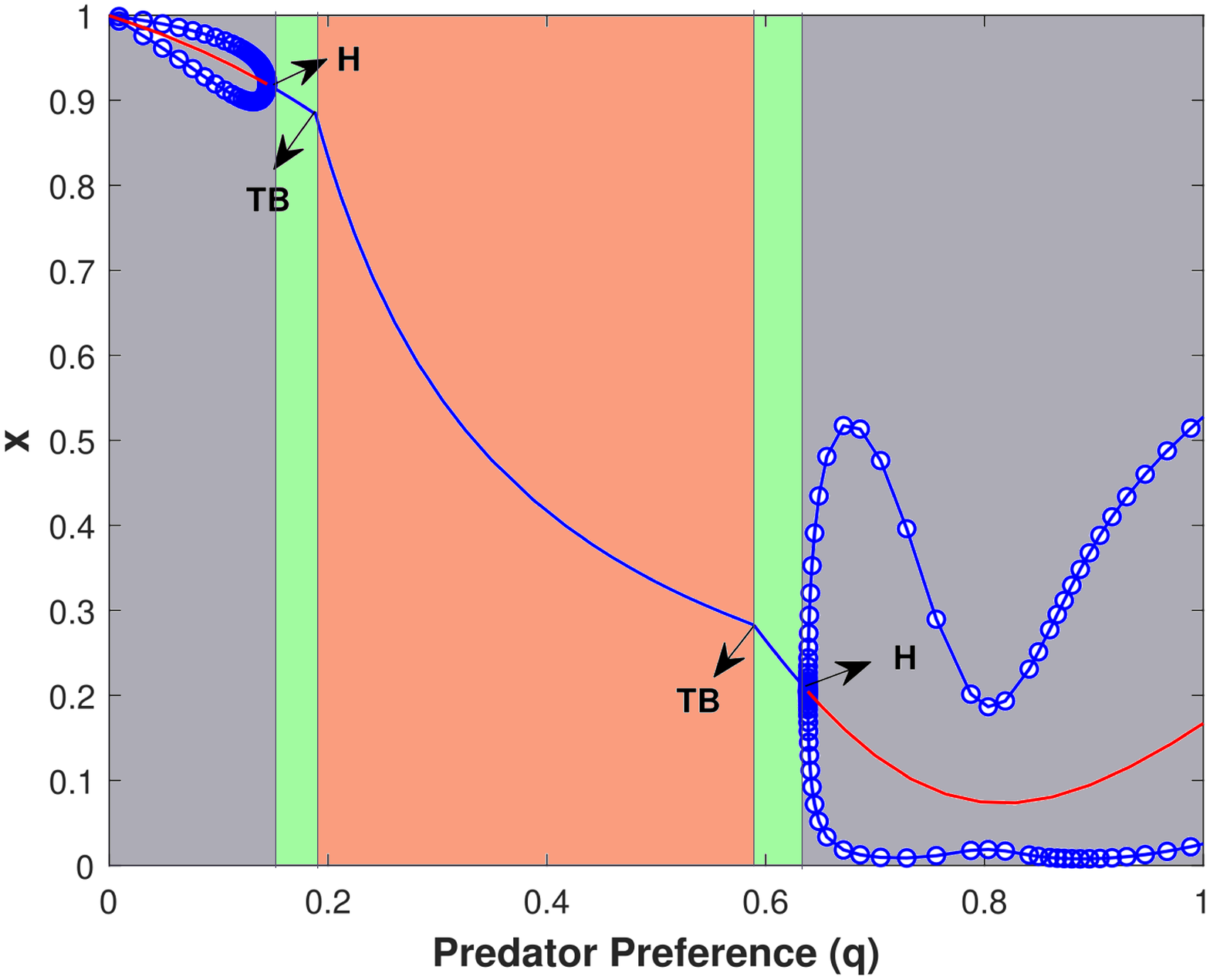}
         \caption{}
         \label{fig:x}
     \end{subfigure}
     \hfill
     \begin{subfigure}[b]{0.33\textwidth}
         \centering
         
         \includegraphics[width=\textwidth]{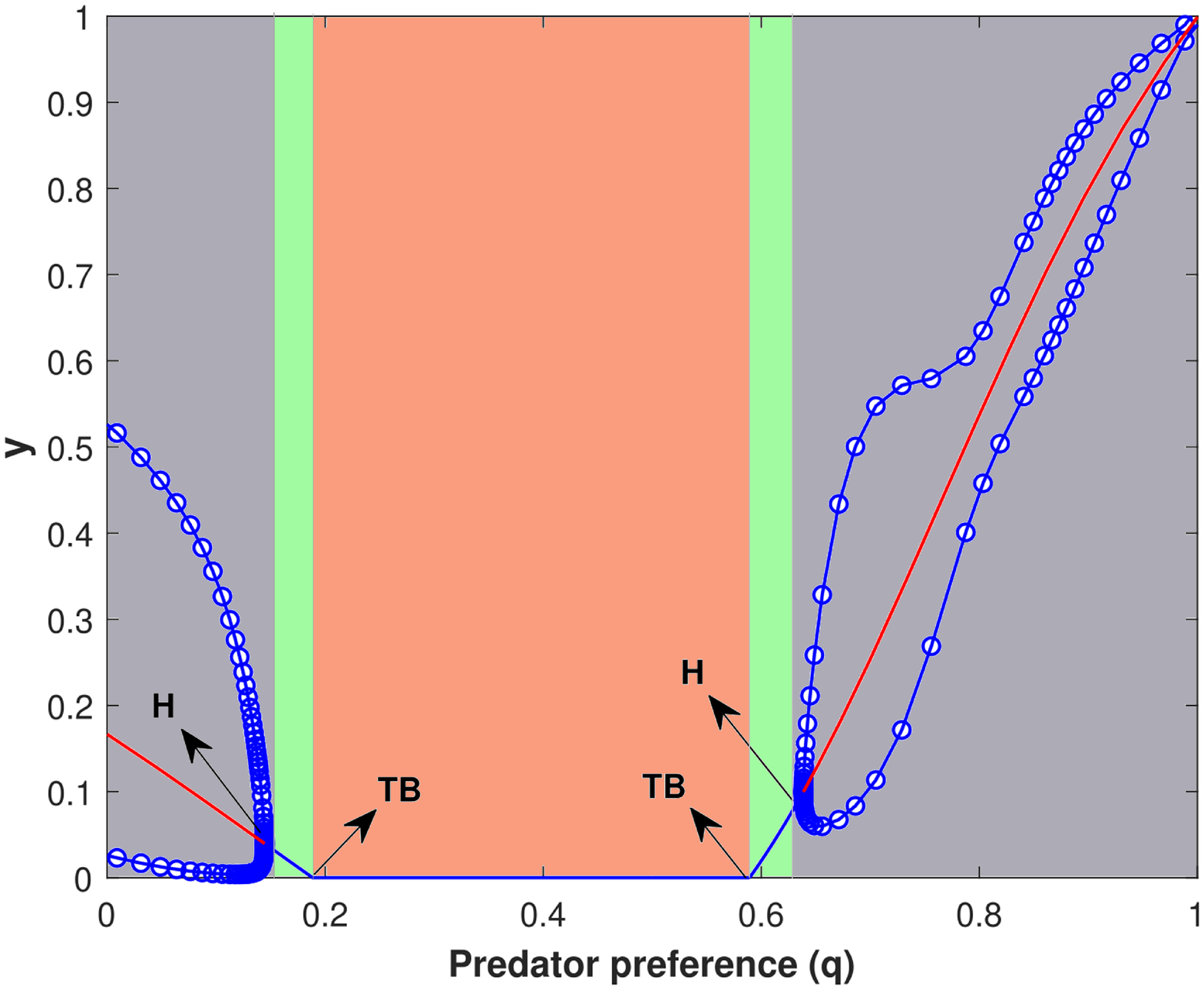}
         \caption{}
         \label{fig:y}
     \end{subfigure}
     \hspace*{-0.6cm}
     \hfill
     \begin{subfigure}[b]{0.329\textwidth}
         \centering
         
         \includegraphics[width=\textwidth]{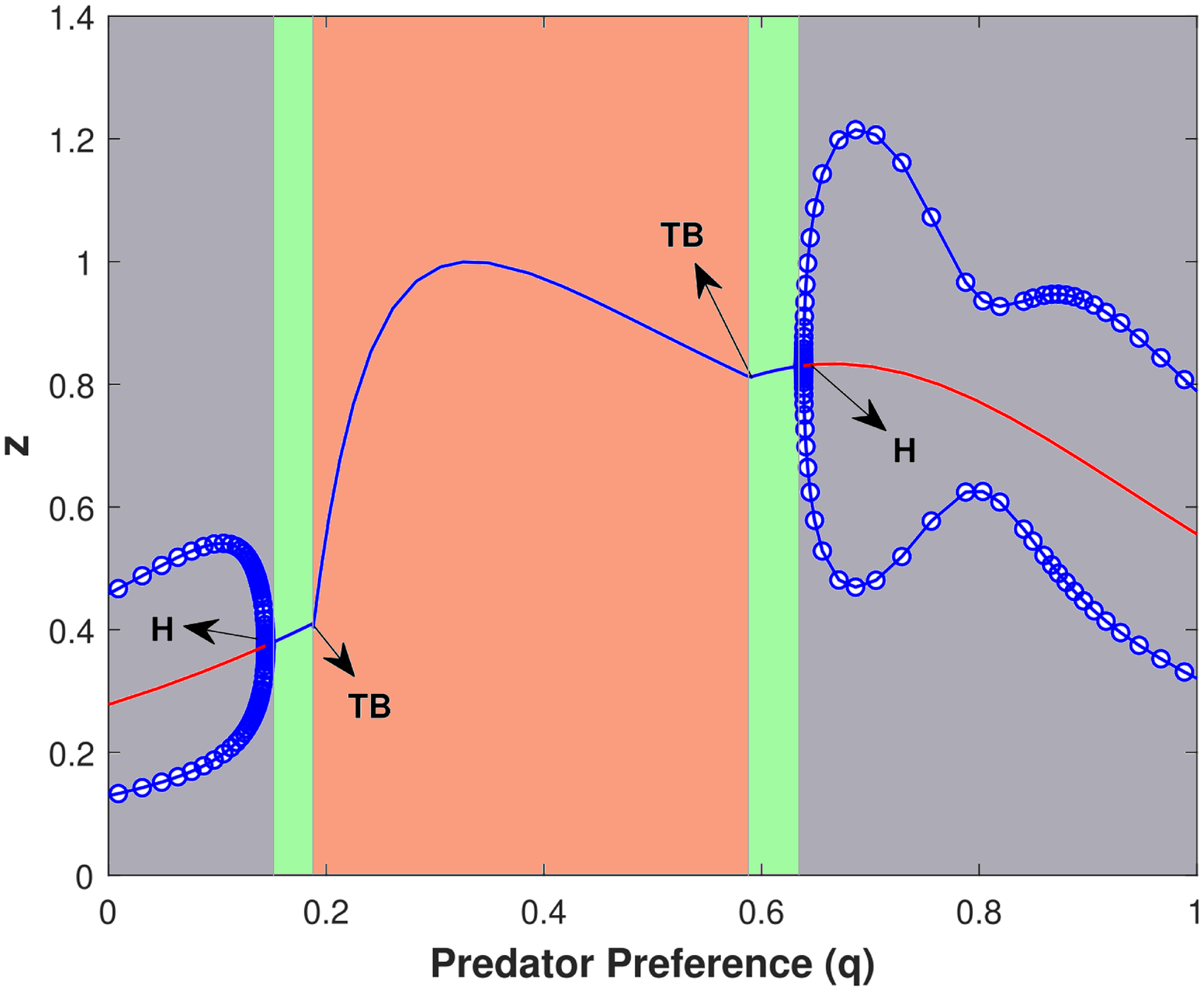}
         \caption{}
         \label{fig:z}
     \end{subfigure}
        \caption{Bifurcation diagram with respect to predation preference (q) with zero contamination (c=0). Blue line denotes stable equilibrium, red line denotes unstable equilibrium and open doted lines indicates maximum, minimum amplitude of the limit cycle. Grey region denotes oscillatory coexistence (OC), green region stands for stable coexistence (SC) and orange region specify uncontaminated prey extinction equilibrium (UPE). The parameter values are $l=1$, $\eta_{1}=0.5$, $\eta_{2}=0.5$, $r=0.5$, $K=1$, $e=0.4$, $\lambda =1.7$, $d=0.1$. }
        \label{fig:One}
\end{figure}

Before investigating the effect of contamination in this three-species food web module, we study our model without contamination to better understand the dynamical consequences when we will incorporate contamination. As predator preference is one of the key parameters of our uncontaminated model, we perform one parameter bifurcation w.r.t predator preference to quantify the qualitative outcomes. Figure \ref{fig:One} depicted that when the predator performs as a specialist, that is, its preference is either very low or very high upon $X$(vice versa for Y) we observe unusable prey-predator dynamics, where prey and predator exhibit a stable limit cycle around the unstable fixed point. Further walking along the average preference from both sides, we find that the dynamics become stable through Hopf bifurcation(H), where three species system coexists in a stable equilibrium and oscillation dies out. When the predator performs as a generalist or its preference is around the average of the preferences, we observe another regime shift through transcritical bifurcation(TB); with this regime shift, the dynamics enter into one prey extinction equilibrium from a coexistence equilibrium(see Figure \ref{fig:One}). This extinction occurs due to apparent competition between prey\citep{holt2017apparent}. The idea of the apparent competition is that if two prey species coexist at equilibrium, the abundance of each is necessarily lower than if it were alone with the predator. Alternatively, if the predator persists and the two prey species cannot coexist, then one prey indirectly excludes the other from the community by sustaining the predator.
\subsection{Prey predator dynamics in contaminated sytem}
\begin{figure}[H]
    \hspace*{-0.9cm} 
    \includegraphics[width=18cm]{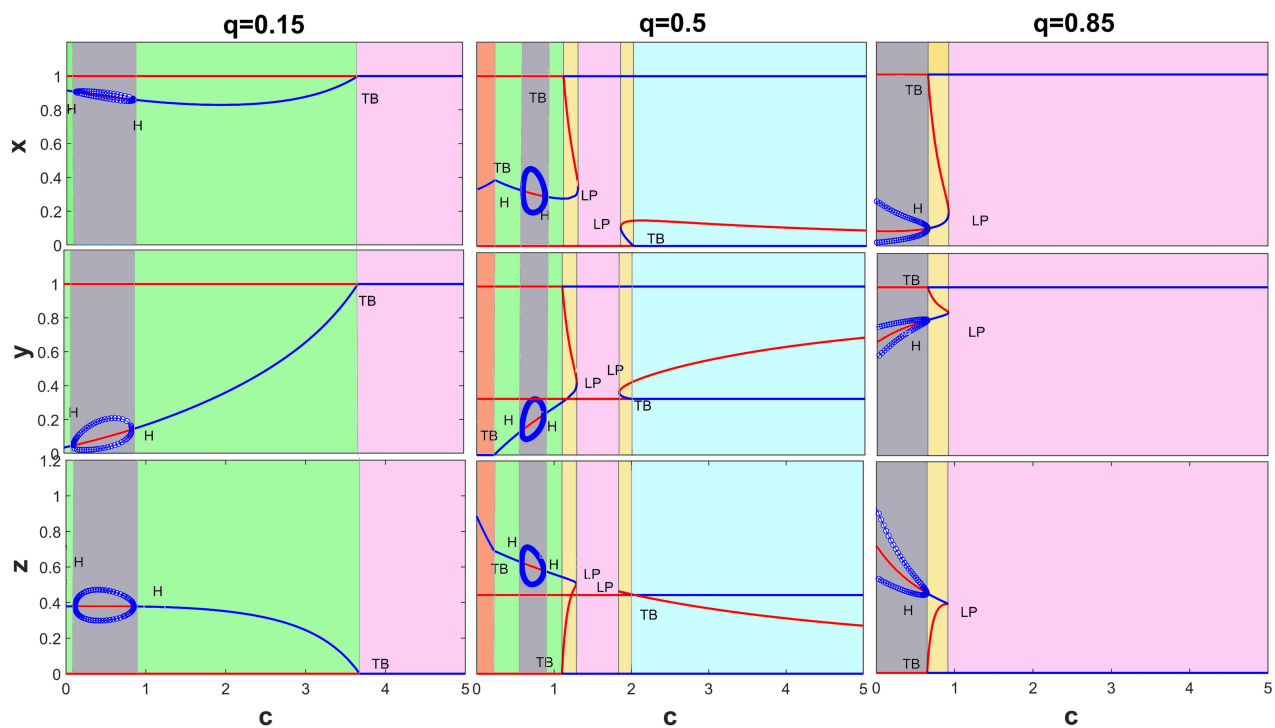}
    \caption{Bifurcation diagram with respect to contamination in prey (c) for different values of predator preference to  contaminated prey (q). Blue line denotes stable equilibrium, red line denotes unstable equilibrium and open doted
lines indicates maximum, minimum amplitudes of the limit cycle. Green region denotes stable coexistence(SC). Orange region denotes uncontaminated prey extinction equilibrium(UPE). Grey region denotes oscillatory coexistence(OC). Yellow region denotes bi-stability between predator extinction equilibrium and coexistence equilibrium(SC, PDE). Pink region denotes predator only extinction equilibrium(PDE) and the skyblue region indicates bi-stability between predator extinction and contaminated prey extinction equilibrium(PDE, CPE). Parameter values are same as in Figure\ref{fig:One}. }
    \label{fig:Two}
\end{figure}

To account the effect of contamination in our model system, we perform a one-parameter bifurcation diagram with contamination in prey. We fixed the predator preference q at three values; low contaminated prey preference(q=0.15), moderate contaminated prey preference (q=0.5), and high contaminated prey preference(q=0.85). The corresponding dynamics are depicted in Figure \ref{fig:Two}.
\newline
At low contaminated prey preference (q=0.15), we observe stable coexistence of three species at low contamination. Moving along the contamination axis, we see the entry and the exit of the stable coexistence state in unstable equilibrium through the initiation and elimination of the hopf bifurcation (H). This phenomenon is also known as the" bubbling effect" in ecological literature due to the shape of the bubble in the
one parameter space and also appear in other ecological studies\citep{alves2017hunting,liz2012hydra}. We also observe that the equilibrium density of the contaminated prey decreases first with the contamination and then increases, while the equilibrium density of the uncontaminated prey consistently increases. With this, we also note that the predator's equilibrium density decreases and become extinct through transcritical bifurcation(TB).

\begin{figure}[H]
    \hspace*{3cm} 
    \includegraphics[width=10cm]{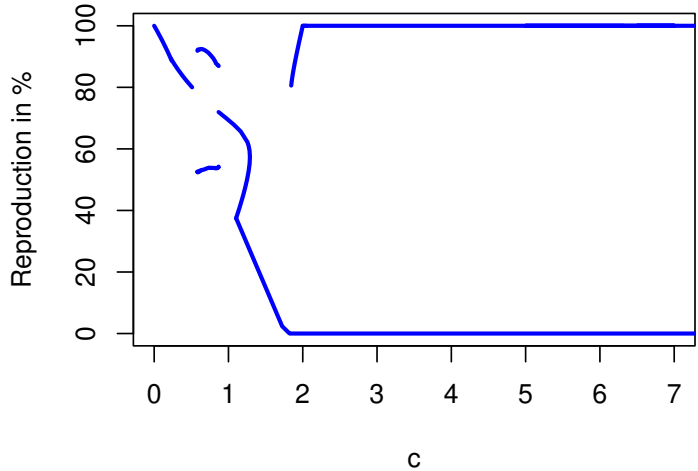}
    \caption{Predator maximum reproduction rate in $\%$=$max(0,1-bc_{p})\times 100$. Where $c_{p}$(contamination in predator) is a function of c(contamination in prey). Parameter values are same as in Figure \ref{fig:One}.  }
    \label{fig:M}
\end{figure}

In Figure \ref{fig:Two}, at the moderate preference for the contaminated prey(q=0.5), moving along the contamination axes-$c$ from zero contamination in prey, we observed equilibrium abundance of the contaminated prey increases. At the same time, the equilibrium abundance of the predator decreases and is also observed by another study \citep{huang2015impact}. That is, contamination affects the predator more than the prey, which releases the contaminated prey from being consumed by the predator, and prey abundance increases. In this case, we observed that uncontaminated prey is in extinction (Figure \ref{fig:Two}). However, after a threshold of the prey contamination, we see the rescue of the uncontaminated prey through transcritical bifurcation. Moving further, we observe that the equilibrium abundance of the contaminated prey and predator decreases while the equilibrium abundance of the uncontaminated prey increases. We also observe the bubbling effect as in the case of low preference for the contaminated prey. Followed by the disappearance of the unstable dynamics, we find bistability between two stable fixed points with one in predator extinction state(see Figure \ref{fig:Two}). Mathematically bistability arises due to saddle-node bifurcation, where the stability of the coexistence equilibrium is lost by colliding with another unstable coexistence equilibrium. The bistability region is from the transcritical point (c=1.103) up to the limit point (c=1.29). In this bi-stability region,  the population will go to the predator extinction equilibrium(PDE) for some initial conditions, while for others, in stable predator existence equilibrium(SC). Ecologically in this region, a considerable perturbation can bring the predator to extinction from the existing state or lead to predator establishment from the extinction state. The threshold density above which the predator can establish the prey's coexistent equilibrium or the critical density below which the predator will go to extinction from the coexistence equilibrium is the allee threshold of the predator \citep{van2005bistability}. 
\newline
Moving further, we find the disappearance of the bistable state through the limit point(LP, c=1.29) and the existence of the predator only extinction equilibrium (PDE)(see Figure \ref{fig:Two}) where prey contamination drives predator to the extinction due to reduction of the reproduction(see Figure \ref{fig:M}). Following the extintion of the predator, both the prey reach to the carrying capacity due to the release from the predation(see Figure \ref{fig:Two}). We observe the region from the second limit point(c=1.84) to the second transcritical point (TB) at(c=2). In this region, two alternative fixed points exist one predator existence and another predator extinction equilibrium. That is, the predator can established to the system in the prey coexistence equilibrium. After the transcritical point, we observe two stable fixed points; in one, contaminated prey exists but the predator is in extinction. While in other, the predator exists, but the contaminant prey is in an extinction state. That is, if the predator is absent in the system and two prey exists, then the predator can established to the system if its density is above some critical density. However, it ultimately brings the contaminant prey to extinction due to apparent competition. On the other hand, if a predator and uncontaminated prey exist in the system, but contaminant prey is absent, contaminant prey can established the system if its density is above a critical density but drives the predator to extinction due to contamination.
\newline
At the high contaminated prey preference (q=0.85) we see the oscillation of three species(OS) at low contamination. Moving further, we observe that the dynamics become stable through hopf bifurcation (H). Followed by the hopf bifurcation, we see the equilibrium density of the contaminated prey and uncontaminated prey increase, and the equilibrium density of the predator decreases until the system goes to the saddle-node bifurcation where two equilibrium points exist, one is three species stable coexistence (SC) and another is predator extinction equilibrium (PDE). 

\subsection{Two Parameter bifurcation diagram}
In order to get insights more to the interplay between predation preference and contamination concentration in prey we perform two parameter bifurcation in bi-parameter space $q-c$. We also interested in the joint impact of contamination intensity in predator reproduction and the contamination in prey, to investigate it we perform two parameter bifurcation analysis in $\lambda -c$ plane. The dynamics of the system in different bi-parameter spaces (simultaneous variation of two parameters) would be interesting and may
reveal rich dynamical behaviors in the system.
\subsubsection{Dynamical interplay between predation preference(q) and contamination in prey (c)}
\begin{figure}[H]
     \hspace*{-2cm}
         \centering
         \includegraphics[width=16.5cm]{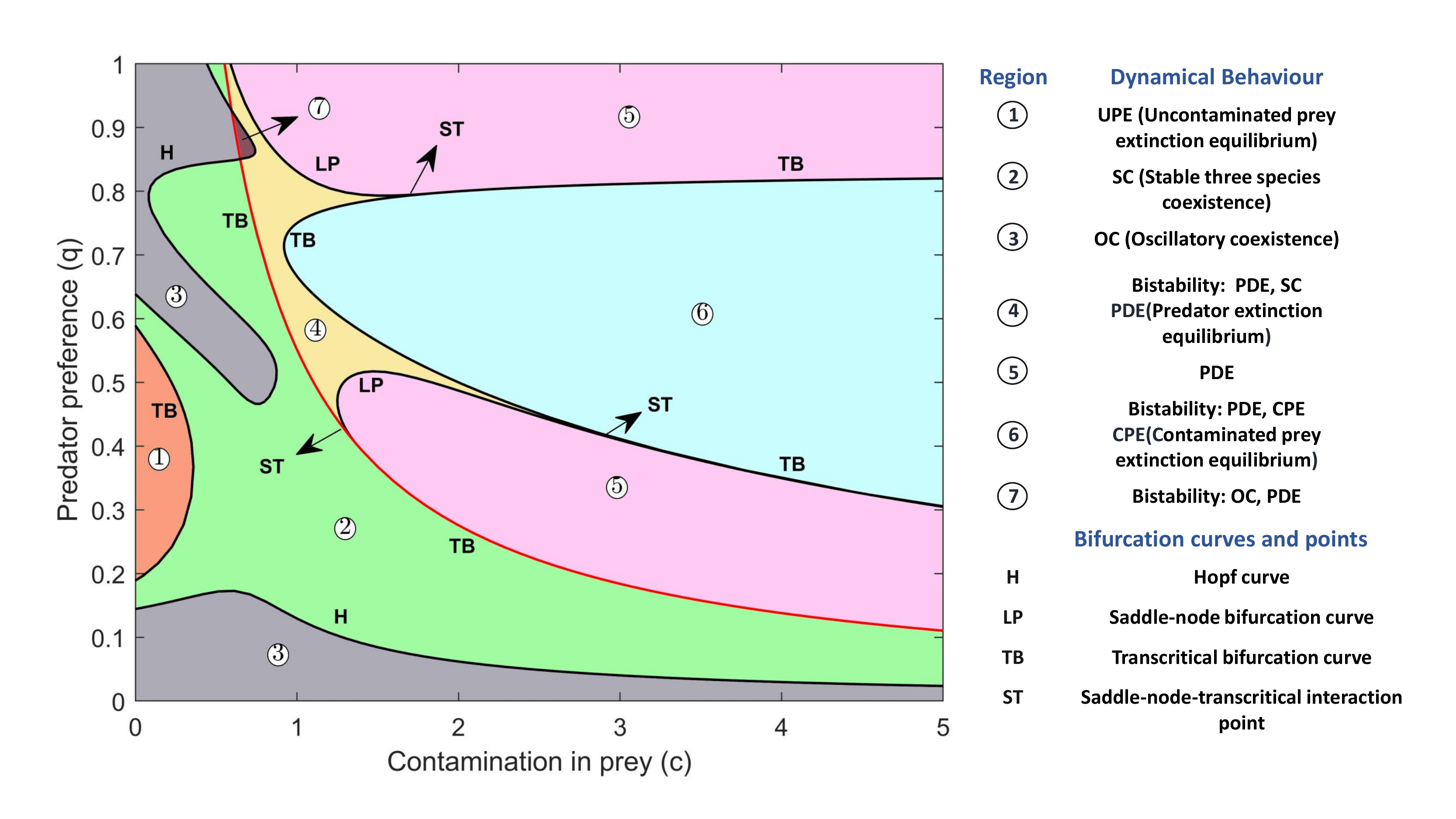}
        \caption{Two parameter bifurcation diagram in $q-c$ space. Red line separates predator invasion region. Left of this line predator always persists in any region and for any initial conditions while in the right side conditionally persists in some region and in others in extintion state. We choose $\lambda$=3 and other parameter values are same as in Figure \ref{fig:One}.  }
        \label{fig:Four}
\end{figure}

In Figure\ref{fig:Four}, the $q-c$ parameter space  is divided into seven distinct regions of different dynamical behaviours and are separated by lines which represents different types of bifurcation curves. Moving along the contamination axis with low contaminated prey preference we observe the dynamics in region \textcircled{3} where three species exhibits oscillatory coexistence (OC) and at high contamination the system enter into region \textcircled{2} where the dynamics become stable through hopf bifurcation. In region \textcircled{1} the uncontaminated prey become extinct and with contamination the dynamics enter into region \textcircled{2} through transcritical bifurcation and the uncontaminated prey came back to the system. Further increasing the contamination, the system move from region \textcircled{2} to region \textcircled{5} for low contaminated prey preference or in region \textcircled{4} at moderate contaminated prey preference. In region \textcircled{5} the contaminated predator become extinct followed by transcritical bifurcation. If the contamination in prey increases further we observe bistability between PDE and CPE in region \textcircled{6}. At high contaminated preference and low or no contamination in prey, we still observe the oscillatory dynamics in region \textcircled{3}, with increasing contamination the dynamics either enter to region \textcircled{7} through transcritial curve or via region \textcircled{2} it fall in region \textcircled{4}. In region \textcircled{7} we see bistability between OC and PDE. At region \textcircled{4} we observe bistability between SC and PDE.
\newline
When contamination of prey is low(c=0.5), moving along the prey preference axis from zero to one we observe the dynamical behaviour fall in region \textcircled{1}, \textcircled{3}. That is, at low contamination contaminated prey preference dose not effect predator persistence. At moderate contamination walking along the prey preference we see the shift of the dynamical behaviour from region \textcircled{2} to region \textcircled{4} or region \textcircled{5} through the transcritical bifurcation curve. This transcritical bifurcation curve is the predator persistence line (red line). In the left hand side of this line the predator always persist but in the right hand side of this line predator dose not persist in some region and conditionally persist in other regions. In this bi-parameter space we also observe Saddle-node Trancritical  interaction point (ST), which have important ecological significance. At this point allee effect despair by colliding the saddle node point (LP) with transcritical point (TB), that is system enter to a smooth extinction state from abrupt extinction state\citep{donohue2018technique,van2019automatic,donohue2020normal}. 
\begin{figure}[H]
     \centering
     \begin{subfigure}[b]{0.45\textwidth}
         \centering
         \includegraphics[width=\textwidth]{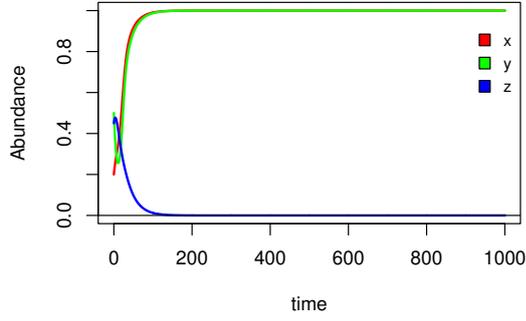}
         \caption{}
         \label{fig:a1}
     \end{subfigure}
     \hfill
     \begin{subfigure}[b]{0.45\textwidth}
         \centering
         \includegraphics[width=\textwidth]{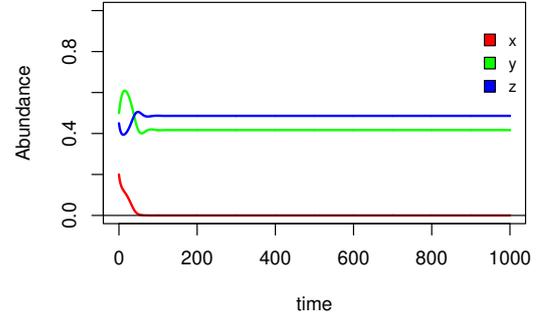}
         \caption{}
         \label{fig:a2}
     \end{subfigure}
     \hfill
     \begin{subfigure}[b]{0.45\textwidth}
         \centering
         \includegraphics[width=\textwidth]{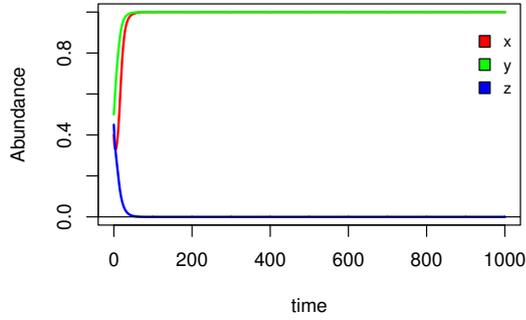}
         \caption{}
         \label{fig:a3}
     \end{subfigure}
     \hfill
     \begin{subfigure}[b]{0.45\textwidth}
         \centering
         \includegraphics[width=\textwidth]{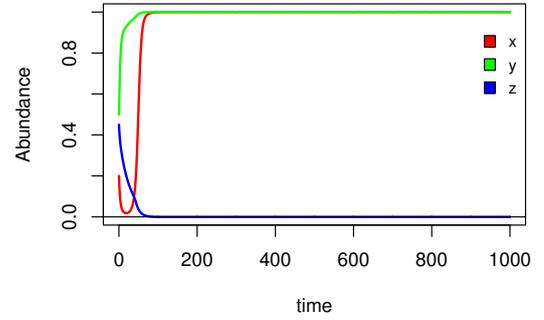}
         \caption{}
         \label{fig:a4}
     \end{subfigure}
     \caption{Time series of abundance at region \textcircled{5} and \textcircled{6}. (a)  c=2.3, q=0.3 with initial condition (0.2,0.5,0.45); (b) c=2.3, q=0.6 with initial condition (0.2,0.5,0.45); (c) c=2.3, q=0.6 with initial condition (0.4,0.5,0.45), (d) c=2.3, q=0.85 with initial condition (0.2,0.5,0.45). Rest of the parameter values are same as in Figure \ref{fig:One}.}
     \label{fig:Timeq_c}
\end{figure}

\subsubsection{Dynamical interplay between intensity of contamination in predator reproduction\texorpdfstring{($\lambda$)} and contamination in prey (c)}
\begin{figure}[H]
     \hspace*{-1cm}
         \centering
         \includegraphics[width=16.5cm]{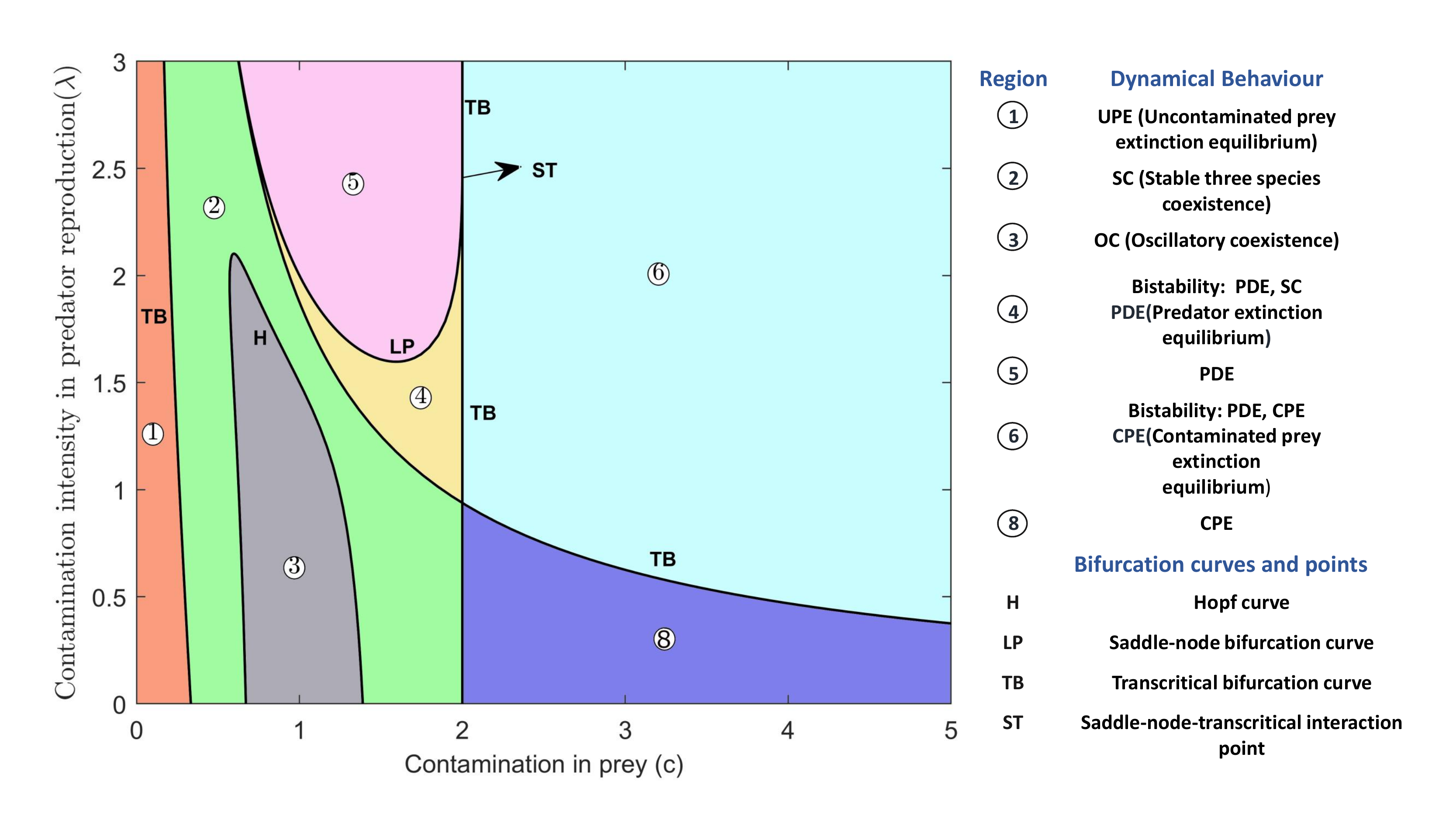}
         \caption{Two parameter bifurcation diagram in $\lambda-c$ space. We chosse the value of q as 0.5 and rest of the parameter values are same as in Figure \ref{fig:One}. }
        \label{fig:Five}
\end{figure}

To demonstrate how the intensity of the contamination in predator reproduction and contamination in prey influences prey-predator dynamics, we carried out a two-parameter bifurcation diagram in Figure\ref{fig:Five}. At low contamination, we observe that uncontaminated prey is in an extinction state and contaminated prey and predator stably exist in region \textcircled{1}. Moving along the contamination axis,  we observe that the dynamics enter into region \textcircled{2} through transcritical bifurcation (TB). In this region, three species coexist in stable equilibrium (SC). We also observe the region \textcircled{3} surrounded by the Hopf curve of region \textcircled{2}, where three species exhibit unstable dynamics. When the effect of the contamination in predator reproduction is moderate, we observe that the dynamics enter from the region \textcircled{2} to the region \textcircled{4} through the Limit point curve. In this region, the model exhibit bi-stable dynamics between predator extinction equilibrium (PDE) and stable coexistence equilibrium (SC). If the severity of the contamination in predator reproduction further increases, the predator can not persist to the system in region \textcircled{5}. 
\newline
At high contamination, if the intensity of the contamination in predator reproduction is low, we observe the extinction of the contaminated prey (CPE) in region \textcircled{8}. Where the region \textcircled{8} is separated from \textcircled{2} by the transcritical curve (TB), on the contrary, increasing the intensity of the contamination in predator reproduction, the dynamics can enter into the bi-stable region \textcircled{6}. In this region, the system's dynamics can reach the predator extinction equilibrium (PDE) for some initial conditions while for others, in the contaminated prey equilibrium (CPE).

\subsection{Robustness}\label{Robustness}
\subsubsection{Robustness for other parameter sets}\label{Robustness1}
\begin{figure}[H]
\hspace*{0cm} 
     \centering
     \begin{subfigure}[b]{0.5\textwidth}
         \centering
         \includegraphics[width=\textwidth]{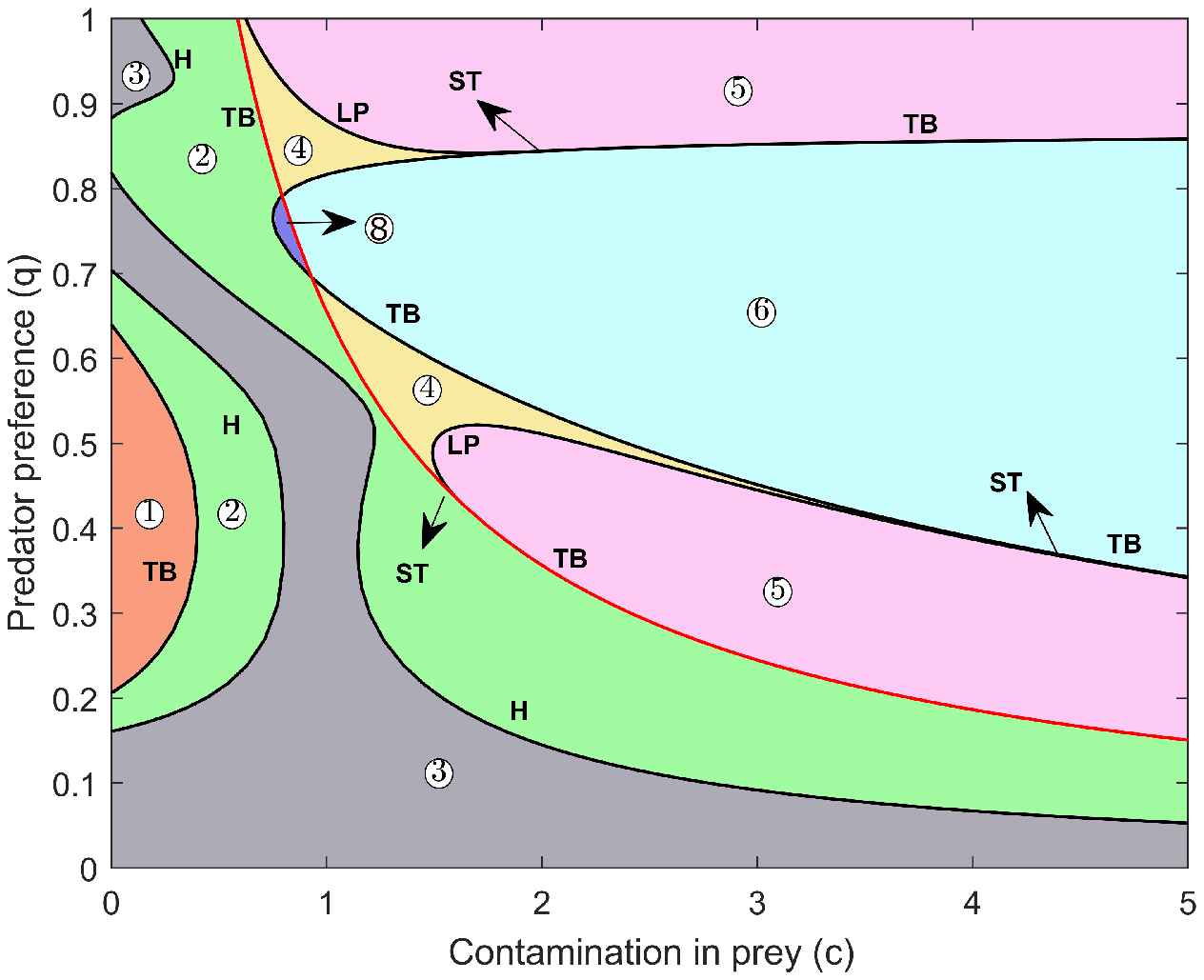}
         \caption{}
         \label{fig:b1}
     \end{subfigure}
     \hspace*{-2cm}
     \hfill
     \begin{subfigure}[b]{0.5\textwidth}
         \centering
         \includegraphics[width=\textwidth]{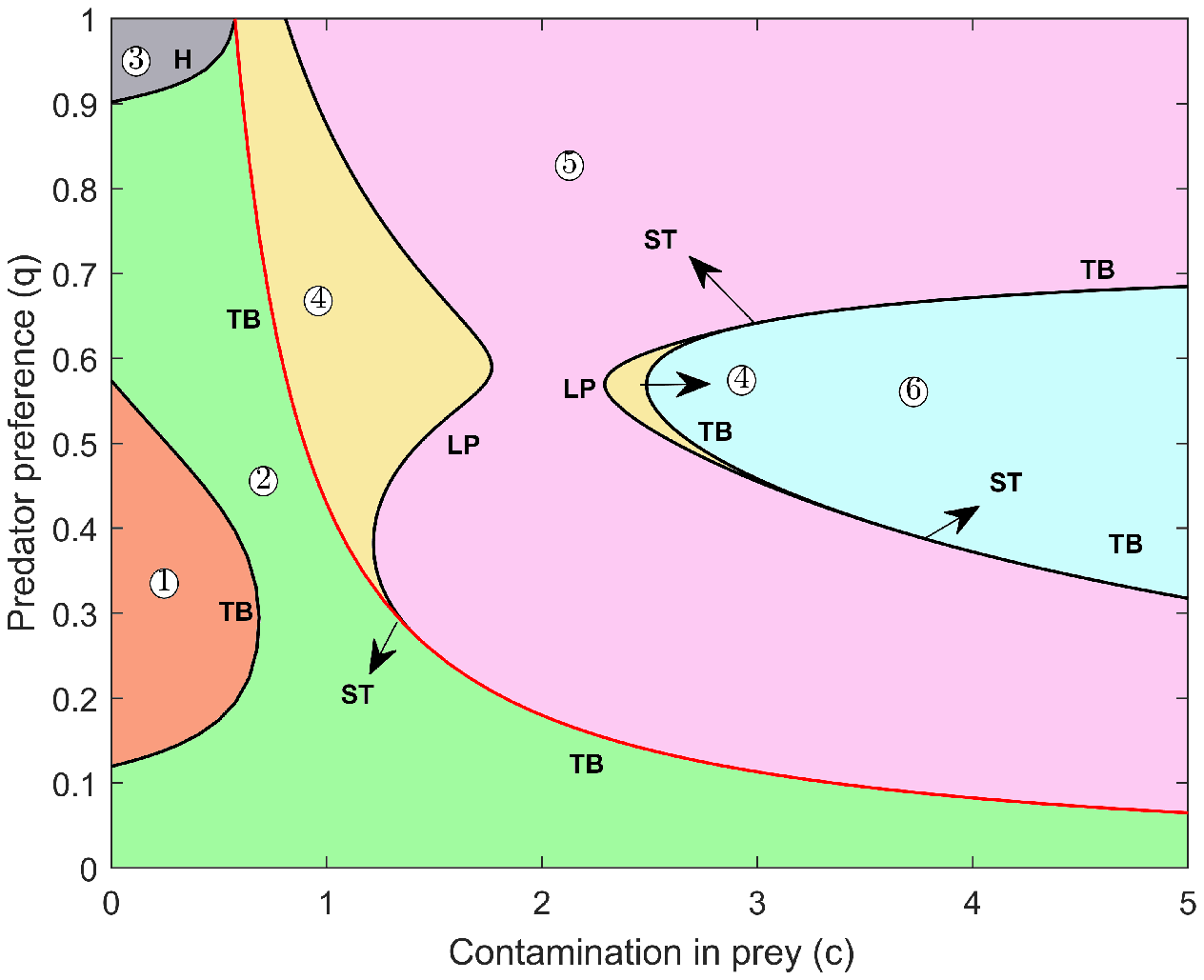}
         \caption{}
         \label{fig:b2}
     \end{subfigure}
     \hspace*{-0.74cm}
     \hfill
     \begin{subfigure}[b]{0.5\textwidth}
         \centering
         \includegraphics[width=\textwidth]{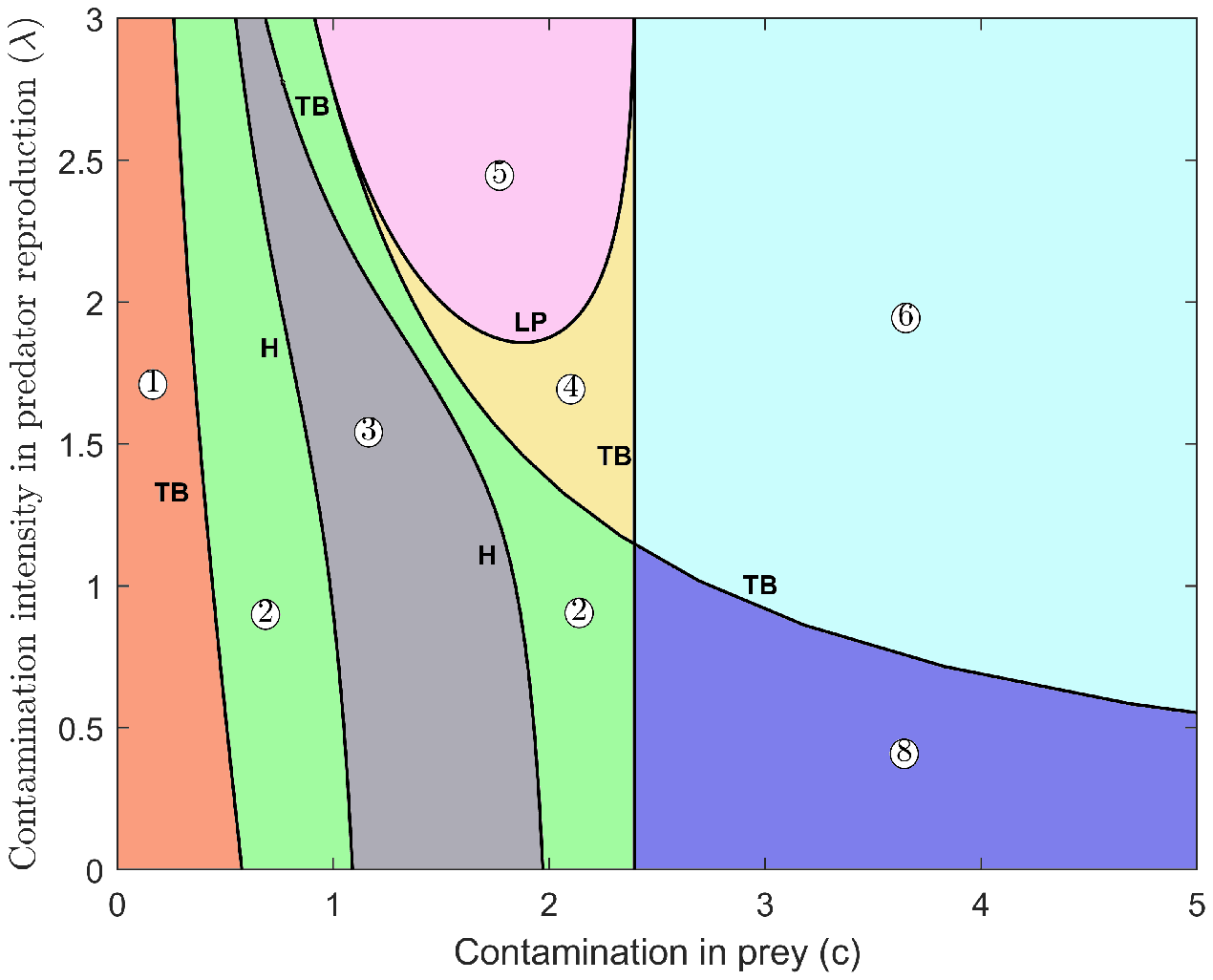}
         \caption{}
         \label{fig:b3}
     \end{subfigure}
     \hspace*{-1.2cm}
     \hfill
     \begin{subfigure}[b]{0.5\textwidth}
         \centering
         \includegraphics[width=\textwidth]{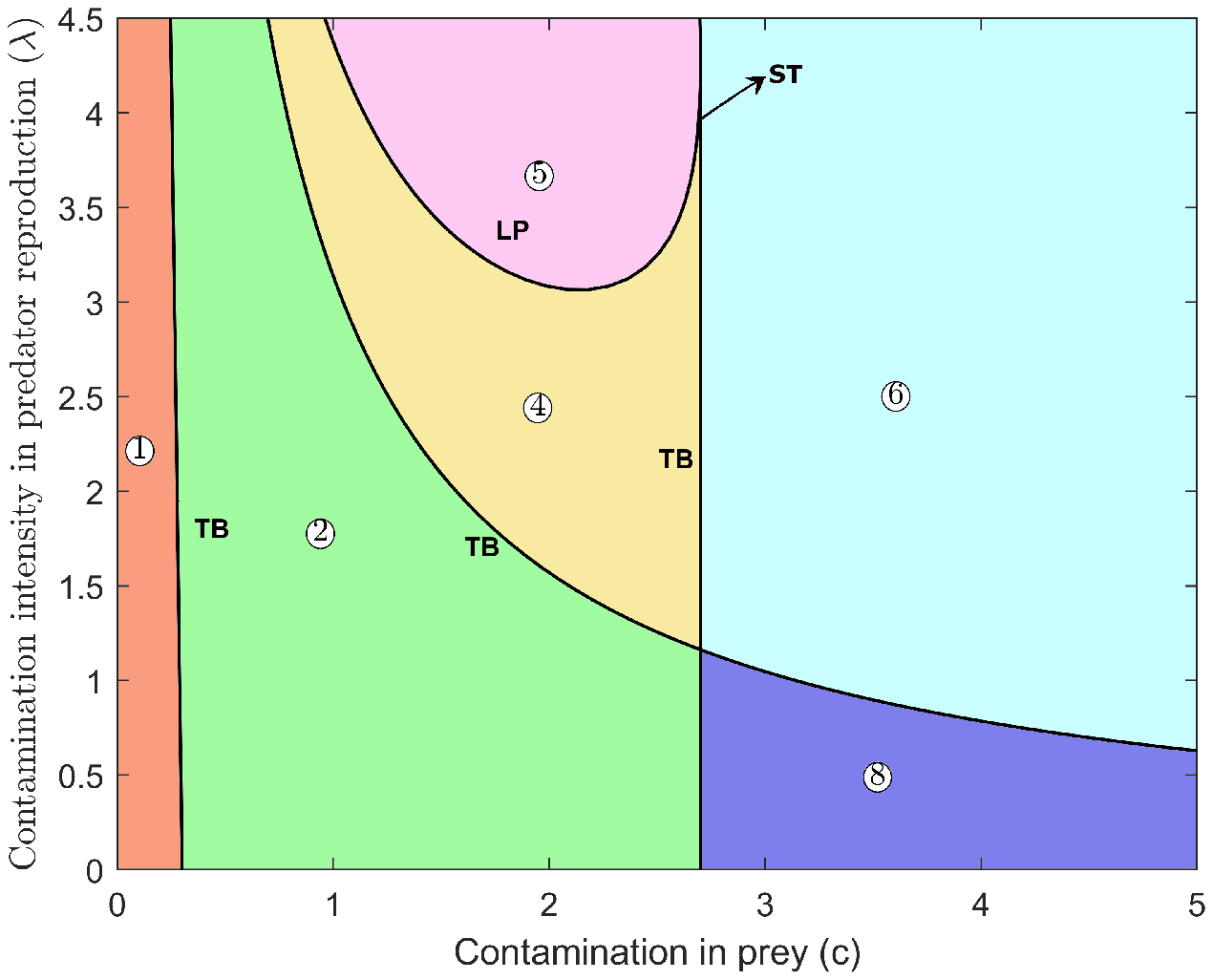}
         \caption{}
         \label{fig:b4}
     \end{subfigure}
        \caption{(a,b) represents two parameter bifurcation diagram with respect to predator preference for contaminated prey (q) and  contamination in prey (c) while (c,d) represents two parameter bifurcation diagram with respect to contamination intensity in predator reproduction ($\lambda$) and  contamination in prey (c). We choose parameter values for (a,b) as $l=0.85$, $\eta_{1}=0.6$, $\eta_{2}=0.6$, $r=0.4$, $K=1.4$, $e=0.6$, $d=0.12$ where for (a) we set  $\lambda =2$, and in (c) we set $q=0.5$. We choose parameter values for (b,d) as $l=0.65$, $\eta_{1}=0.7$, $\eta_{2}=0.7$, $r=0.6$, $K=0.4$, $e=0.85$, $d=0.075$ where for (b) we set $\lambda =3.5$, and in (d) we set $q=0.5$. The dynamical behaviour and bifurcation curves have their usual meaning as mentioned in Figure \ref{fig:Four} and \ref{fig:Five}.}
        \label{fig:Six}
\end{figure}

Along with the above parameter set, we consider two additional parameter sets to ensure that the dynamics we have found are not a particular case. The dynamics in new parameter sets depicted in Figure \ref{fig:Six} The qualitative behavior of this species interaction module is almost identical, except some new regions appear and some previous regions disappear in these new parameter combinations. For instance the three species oscillation region \textcircled{3} expand in \ref{fig:Six}(a,c) while in \ref{fig:Six}(b,d) is it become more narrow or completely absent as compared to Figure \ref{fig:Four},\ref{fig:Five}. Further region \textcircled{7} where we see the bi-stability between PDE and OC is completely absent in Figure \ref{fig:Six}(a,b).
\subsubsection{Robustness for alternative functional form of prey reproduction and incorporation of contamination induced prey mortality}\label{Robustness2}
\begin{figure}[H]
\hspace*{0cm} 
     \centering
     \begin{subfigure}[b]{0.5\textwidth}
         \centering
         \includegraphics[width=\textwidth]{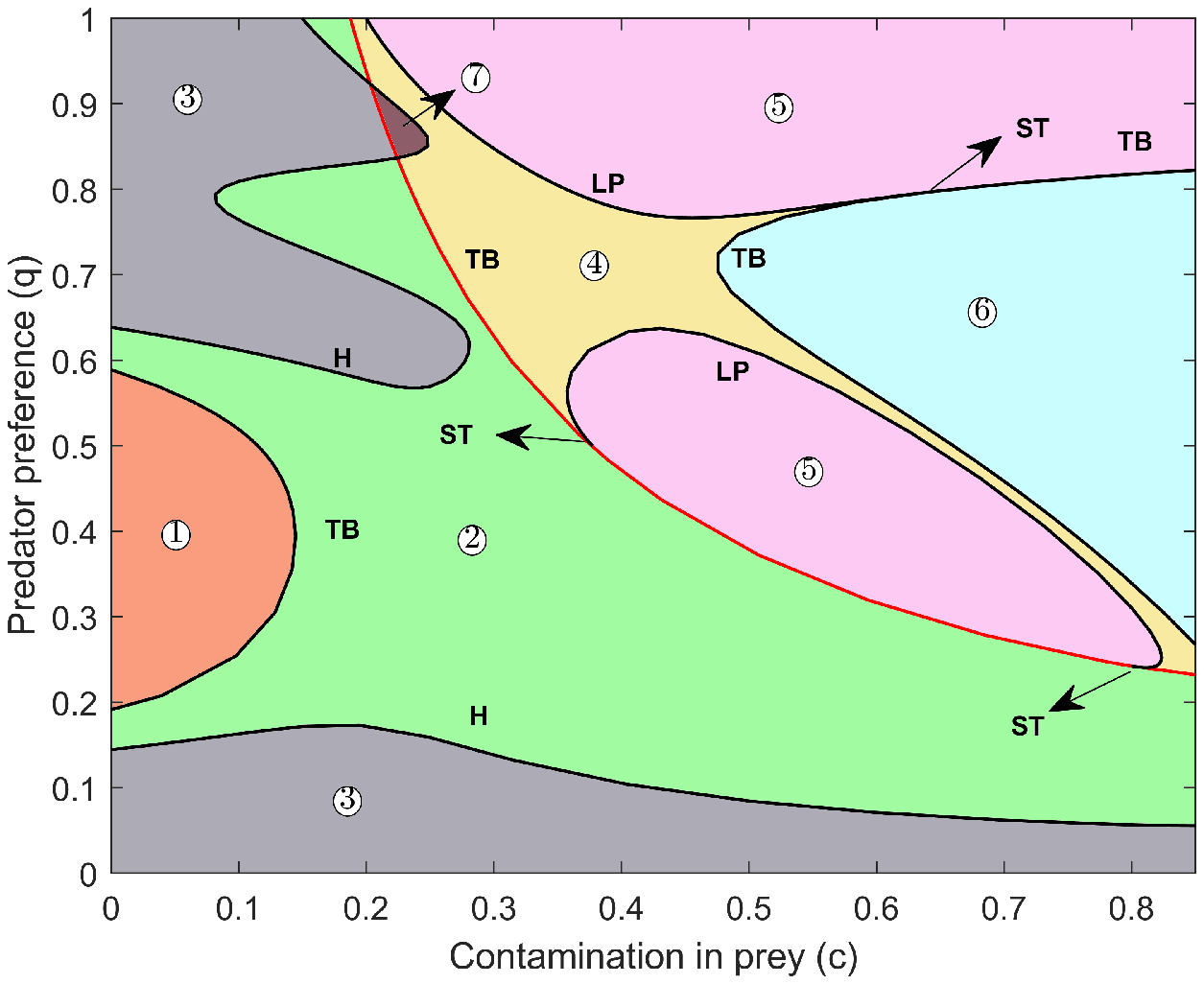}
         \caption{}
         \label{fig:c1}
     \end{subfigure}
     \hspace*{-2cm}
     \hfill
     \begin{subfigure}[b]{0.5\textwidth}
         \centering
         \includegraphics[width=\textwidth]{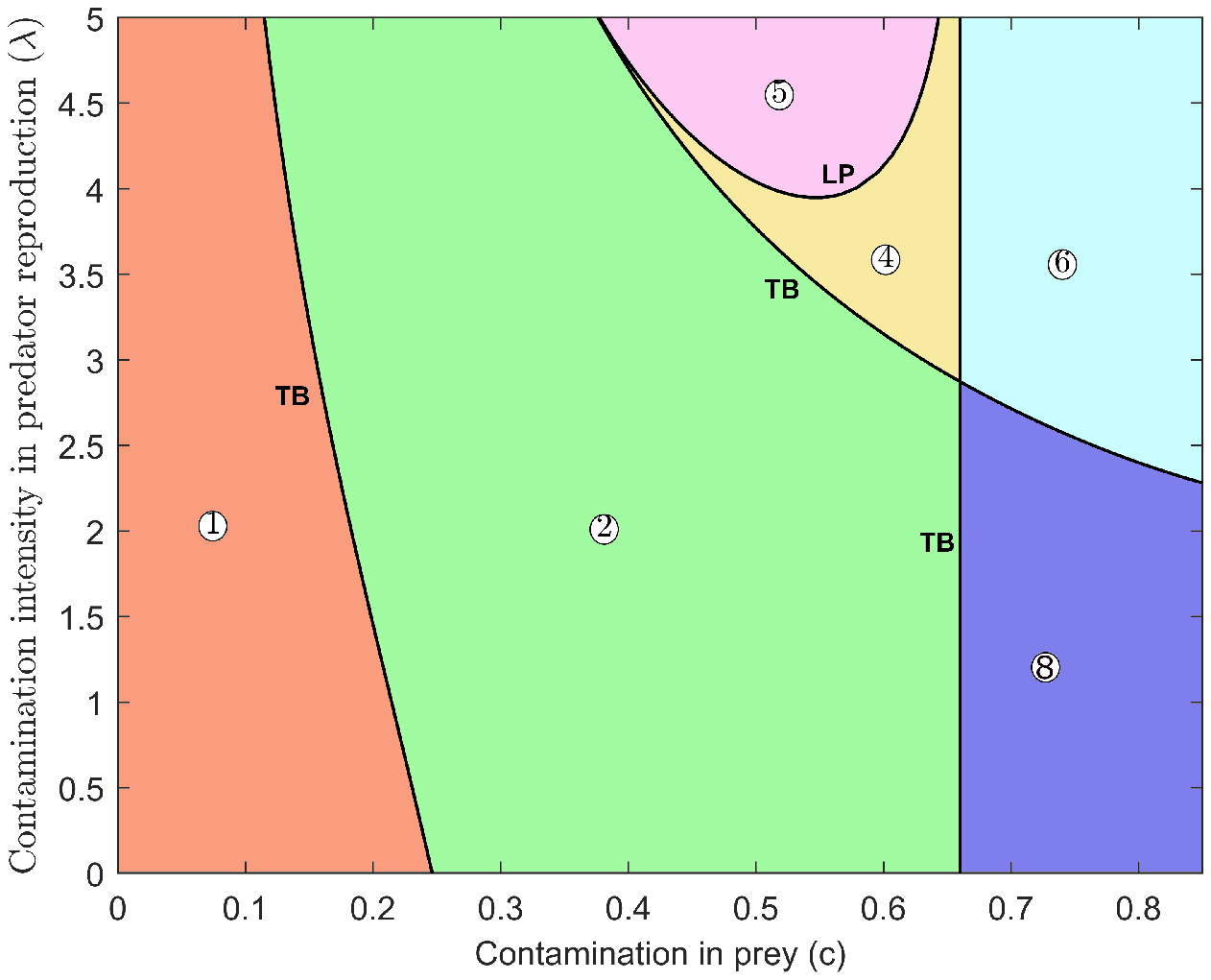}
         \caption{}
         \label{fig:c2}
     \end{subfigure}
        \caption{(a) represents two parameter bifurcation diagram with respect to predator preference for contaminated prey (q) and  contamination in prey (c) while (b) represents two parameter bifurcation diagram with respect to contamination intensity in predator reproduction ($\lambda$) and  contamination in prey (c). We choose parameter values for as $l=1$, $\eta_{1}=0.5$, $\eta_{2}=0.5$, $r=0.5$, $K=1$, $e=0.4$, $\lambda =5$, $d=0.1$, $m_{1}$=0.01 in (a) we set $\lambda$=5 and in (b) we set $q=0.5$. The dynamical behaviour and bifurcation curve have their usual meaning as mentioned in Figure \ref{fig:Four} and \ref{fig:Five}.}
        \label{fig:three graphs 3}
\end{figure}

Our main analysis assumes a nonlinear functional form for the reproduction of contaminated prey with contamination. We relax this assumption by replacing the function $\frac{1}{1+c}$ with the linear response of contamination $max(0,1-c)$ in prey reproduction. Further, in the primary analysis, we did not assume any contaminated-related death in the contaminated prey. We add the term $m_{1}cx$ in the equation (2.13) to incorporate contamination-related death of the prey. Where $m_{1}$ is the effect of toxin on the mortality, c is the contamination in prey and $x$ is the contaminated prey density. Figure \ref{fig:three graphs 3} is the representation of the of the Figure \ref{fig:Four}, Figure \ref{fig:Five} for the parameter set given in the caption of Figure \ref{fig:three graphs 3}. We can observe that the qualitative dynamical behavior of the system more or less same with previous assumption.

\newpage
\section{Discussion}\label{Discussion}
Methylmercury (MeHg) was previously thought only to threaten aquatic biota and consumers of aquatic organisms, as aquatic environments provide ideal conditions for the conversion of inorganic mercury to MeHg\citep{tweedy2013effects}. However, recent studies have found elevated concentrations of MeHg in terrestrial consumers, which seeks special attention in conservation management due to the worldwide decline of insectivores\citep{manning2021conservation}. The main reason behind the transfer of aquatic contamination to terrestrial consumers is the predation of emergent aquatic insects \citep{jackson2021differential}. It is well known that aquatic prey can negatively affect terrestrial preys\citep{polis1997toward,baxter2005tangled,murakami2002indirect}through the increase of predator density. Furthermore, contamination can also drive insectivores and aquatic prey to the edge of extinction, which may also depend on predation. So, it is necessary to understand how contamination drives these species' dynamics to quantify the risk of contamination more preciously. In connection to this, in this work, we have examined the effect of harmful and beneficial prey associated with contamination on predator persistence. In addition, we also investigate the combined impacts of contamination and indirect competition in the prey-predator dynamics and how ecological and environmental factors shape community stability and species coexistence mechanisms. To do this, we connect the classical apparent competition model\citep{abrams1998apparent,faria2009interplay}  with the contamination in one of the prey. Where contamination decreases the reproduction rate of the prey, predator and the predator intake both prey with some preferences. We assume that the predator did not uptake contamination through the environment because here we consider the situation where aquatic insects transport contamination to the terrestrial predator \citep{kraus2019contaminants,schiesari2018metacommunities}. We demonstrate that this ecological model has a rich repertoire of dynamical behaviors, including different species coexistence equilibrium, alternative stable states, and stable limit cycles.
\subsection{Indirect effect of contamination in uncontaminated prey}
We first consider the case where there is no contamination in the system, and prey-predator dynamics depend upon the predator's preference. We observe the vulnerable region in the one-parameter space where uncontaminated prey can not persist in the system due to apparent competition(see Figure \ref{fig:One}). Such species extinction due to indirect competition between prey has previously occurred in different studies \citep{faria2009interplay,chakraborty2014harmful}.  Further, we investigate how contamination affects the system when uncontaminated prey is in extinction (see Figure \ref{fig:Two}). We point out that a small amount of contamination is necessary for the three species' coexistence when predation preference is relatively higher toward uncontaminated prey. It occurs because contamination reduces the positive bottom-up effect of contaminated prey on the predator and decreases the predator population. As a result, top-down control on the uncontaminated prey decreases, which initiates its persistence. Thus a small amount of contamination can rescue the alternative prey in indirect competition; in direct competition, such effect is also investigated by \cite{shan2019direct}. Note that, such a strictly positive effect of contamination on alternative uncontaminated prey can be observed if predation preference is high enough towards the alternative prey than the focal prey (see region \textcircled{1} in Figure\ref{fig:Four}). For instance, \cite{paetzold2011environmental} did not find any significant positive effect of aquatic contamination on terrestrial prey. One reason behind it is that the predator prefers aquatic prey over terrestrial prey. We expect a positive effect of contamination in uncontaminated terrestrial prey where the uncontaminated prey is more depressed by predation due to aquatic subsidiary \citep{murakami2002indirect}.
\subsection{contamination can destabilize dynamics and effect of contamination in contaminated prey, predator}
Previous studies reported that contamination stabilizes prey-predator dynamics when prey and predator are exposed simultaneously to a toxin or predator intake toxins through prey only\citep{huang2015impact,baudrot2018effects}. In our study, we also observe that a high level of contamination can stabilize the system. However, when contamination is low on the prey, and the predator has a low or moderate preference for contaminated prey, in that case, contamination can also destabilize prey-predator dynamics(see Figure \ref{fig:Two}, q=0.15,0.5). We also observe that increasing level of contamination drives the predator to catastrophic extinction as reported by other studies \cite{baudrot2018effects, huang2015impact}. Such an abrupt change involves bi-stability,  a slight change in environmental condition can drive the system to an alternative attractor, but to return it to the original state, a more extensive improvement of the environmental condition requires\citep {scheffer1993alternative}.
Interestingly, when predator consume both preys with equal preference (q=0.5), further increasing the level of contamination generate an alternative stable state at which the predator can persist with its two prey (right yellow region in Figure\ref{fig:Two}, q=0.5). Note that a higher level of contamination always leads to the predator's extinction from its coexistence state due to lower reproduction (see Figure \ref{fig:M}). However, the predator can establish the two prey coexistence equilibrium if its initial density is above some critical threshold and decrease the density of the contaminated prey at such a critical level that it does not hamper the predator persistence. If the prey's contamination level increases further, predator and contaminated prey cannot coexist due to bi-stability. That is, depending upon the initial condition, either contaminated prey or the predator exists with uncontaminated prey in a stable equilibrium. Previous studies highlight that prey can benefit from contamination as it decreases the predator\citep{huang2015impact,prosnier2015modeling}. In our study, we observe that in the presence of alternative uncontaminated prey, the contaminated prey can go to extinction, but the predator can persist in an alternative equilibrium. With this prey extinction, the predator gets its potential reproduction rate with the limited resource (see Figure \ref{fig:M} upper curve).  
\subsection{Role of predation preference and sensitivity of predator to contamination in species persistance}
Predator preference for contaminated and uncontaminated prey can affect its persistence which can be understood by Figure \ref{fig:Four}. In a more polluted environment, more preference for uncontaminated prey is always beneficial to the predator as it prompts its persistence from any positive initial condition; the regions to the left side of the red line exhibit such behavior in our model. Although predators can exist within the system under more polluted prey preference in region \textcircled{6} and \textcircled{4}, these are bi-stable regions that do not guarantee predator persistence. It is interesting
to note that, the region \textcircled{6} is bounded below by region \textcircled{5}, which indicates that if predation preference for contaminated prey drops below a threshold than the contaminated prey can invade the system and drives the predator to extinction. The fact is, if a predator is initially abundant in the system, then a minimum predation pressure require to inhabit the establishment of the contaminated prey; otherwise, contaminated prey establish to the system and ultimately drives predator to the extinction through the trophic transfer of the contaminates. Note that, this predation pressure can not always inhabit contaminated prey establishment,  if the initial density of the contaminated prey is relatively higher then contaminated prey ultimately establish to the system.  Also note that region \textcircled{6} also bounded above by region \textcircled{5}, that implies if predation preference for contaminated prey high enough then predator can not exist due to the higher flow of contamination. Figure \ref{fig:Five} unveil that if the effect of contamination in predator reproduction is low, then the existence of the contaminated prey is more hampered than the predator (region \textcircled{8}).  As we assume that there is no contamination-related mortality of the contaminated prey, we can intuitively say this is the combined effect of two mechanisms. First, contamination limits the prey's growth by reducing their reproduction rate thus decreases the flow of toxicity in per-capita predator population. On the other hand, uncontaminated prey maintains predator density; as a result, the top-down pressure on the contaminated prey increases and becomes extinct. Alternatively, we can say this occurs due to the combined effect of contamination and apparent competition\citep{holt2017apparent}. Further, we observe that predator can not survive at a moderate level of contamination when the intensity of the contamination in predator reproduction is high \textcircled{5}. However, note that increasing further contamination predator can return to the system in region \textcircled{6}. Intuitively, this occurs because at a moderate level of contamination prey growth does not fall much compared to a high level of contamination; thus, a sufficient amount of contamination can flow to the predator and become a cause of predator extinction. However, if the contamination increases in the prey further and predator is initially abundant to the system, then the prey suffer from extinction risk due to lower reproduction and high predation pressure, as a result the predator can persist in the system.      
\subsection{Modelling caveats and future direction}
Our model includes a number of simplifications; for instance, our study did not incorporate any stage structure of the contaminated prey, although, in reality, aquatic insects are in stage structure. Stage structures are known to affect the prey-predator dynamics in various ways \citep{lindmark2019size}. Incorporation of stage structure in this model requires proper knowledge about contamination's effect in different life stages, like the effect of contamination in metamorphosis and stage-specific mortality\citep{wesner2014metamorphosis}. Further, it will enhance the complexity of the model and parallelly the dynamics. So we restrict our model without any stage structure. Future studies can incorporate it, which may create additional results that are more realistic in the present scenario. Also, our model merges all the contaminated, uncontaminated prey in single variables. However, in reality, a predator can consume more than two prey with specified preferences. In our model, we did not highlight this situation.
Moreover, contamination-induced behavioral change is another limitation of our study\citep{saaristo2018direct}. We believe that incorporating these things does not alter our main conclusion. We also assume that terrestrial herbivore prey is totally uncontaminated; a low amount of contamination may present in terrestrial herbivores, which may unlikely affect their growth and the insectivores significantly. For instance, \cite{ortega2019relationship} find that terrestrial preys of spiders contain around nineteen times less contamination as in aquatic prey and also estimated methylmercury in some terrestrial insects is zero. Further \cite{jackson2021differential} also found a significant difference of contamination in terrestrial herbivores and emergent aquatic insects, and some terrestrial insects contain a negligible amount of methylmercury.
\newline
In summary, our focus was to combine the effect of direct and indirect species interactions, such as predation and apparent competition, and their response to contamination in species dynamics. Our model highlights the significant possibilities through which contamination can affect this type of prey-predator dynamics. For instance, our study predicts that at higher contamination, contaminated prey can alone be the victim of contamination if there is an alternative prey in the system. Our results also predict that the negative effect of aquatic prey on terrestrial prey can be diminished if the aquatic ecosystem holds a small amount of contamination. Further, a predator can sustain in response to highly contaminated prey under more uncontaminated prey preference. Overall, this study may help to understand the ecotoxicological process and its consequence in community dynamics involving direct and indirect species interactions.
\newline
\section{Appendix A. Positivity and boundedness of the solution }\label{Appendix A}
\subsection{Appendix A.1. Existence, uniqueness and positive invariance}
The right hand side of the system (2.13-2.15) is continuously differentiable and locally Lipschitz in the first quadrant which implies the existence and uniqueness of
solutions for the system in $R_{+}^{3}$. For positive invariance we rewrite the system as:
\begin{equation*}
    \frac{dX}{d\tau}=F(X)
\end{equation*}
Where $X=[x,y,z]^T \in R_{+}^{3}$ and $F(X)=[F_{1}(X),F_{2}(X),F_{3}(X)]^{T}$. The solutions of the system remain in the first quadrant for any non-negative initial condition for all $\tau\geq 0$, since $F_{i}(X)|_{X_{i}=0} \geq 0$ for all $X_{i}=0$ where $i=1,2,3$

\subsection{Appendix A.2. Boundedness}
Let $B(t)=x(t)+y(t)+z(t)$ and differentiating $B$ once yields
\begin{equation*}
\begin{split}
    \frac{dB}{dt}&
    =\frac{x}{1+c}(1-x)-\frac{lqxz}{\eta+qx+(1-q)y}+ry(1-y/K)
    -\frac{l(1-q)yz}{\eta+qx+(1-q)y}\\
    &+el max(\ 0, 1-\frac{\lambda ql xc}{\eta+qx+(1-q)y} )\ (\ \frac{qxz+(1-q)yz}{\eta+qx+(1-q)y} )\ -dz\\
    &=\frac{x}{1+c}(1-x)+ry(1-y/K)-l(1-e max(\ 0, 1-\frac{\lambda ql xc}{\eta+qx+(1-q)y} ))\ (\ \frac{qxz+(1-q)yz}{\eta+qx+(1-q)y} )\ -dz
    \end{split}
\end{equation*}
Since $\frac{1}{1+c}\leq1$, $1- emax(\ 0, 1-\frac{\lambda ql xc}{\eta+qx+(1-q)y} ) \geq 0$
\begin{equation*}
    \frac{dB}{dt} \leq x(1-x)+ry(1-y/K) -dz
\end{equation*}
for any arbitary positive real number N we get,
\begin{equation*}
    \frac{dB}{dt}+NB \leq x(1-x)+ry(1-y/K)-dz+N(x+y+z)
\end{equation*}
Taking $N\leq d$ we get,
\begin{equation*}
    \frac{dB}{dt}+NB \leq x(1-x+N)+y(r-yr/K+N)
\end{equation*}
The maximum values of $x(1-x+N), y(r-yr/K+N)$ are $\frac{(N+1)^{2}}{4}, \frac{K(r+N)^{2}}{4r}$ respectively. By letting $A=\frac{(N+1)^{2}}{4}+\frac{K(r+N)^{2}}{4r}$ we get
\begin{equation*}
     \frac{dB}{dt}+NB \leq A
\end{equation*}
By diffrential inequality
\begin{equation*}
    0< B(x,y,z)\leq \frac{A(1-exp(-Nt))}{N}+B(x(0),y(0),z(0))exp(-Nt)
\end{equation*}
So for large values of t we have $0 \leq B\leq \frac{A}{N}$. Hence the solution of the system are bounded in the positive quadrant.
\section{Appendix.B. Local stability analysis of the equilibrium points}
Our model exhibit seven possible equilibrium points $E_{0}=(0,0,0)$, $E_{1}=(1,0,0)$, $E_{2}=(0,K,0)$, $E_{3}=(1,K,0)$, $E_{4}=(x_{4},0,z_{4})$, $E_{5}=(0,y_{5},z_{5})$, $E_{6}=(x_{6},y_{6},z_{6})$. The jacobian matrix at any equilibrium point $(x^{*},y^{*},z^{*})$ is:
\begin{equation*}
J=
 \begin{bmatrix}
J_{11} & J_{12} & J_{13}\\
J_{21} & J_{22} & J_{23}\\
J_{31} & J_{32} & J_{33}
\end{bmatrix} 
\end{equation*}
Where $J_{11}=\frac{\partial dx}{\partial x dt}|_{(x^{*},y^{*},z^{*})}$, $J_{12}=\frac{\partial dx}{\partial y dt}|_{(x^{*},y^{*},z^{*})}$, $J_{13}=\frac{\partial dx}{\partial z dt}|_{(x^{*},y^{*},z^{*})}$, $J_{21}=\frac{\partial dy}{\partial x dt}|_{(x^{*},y^{*},z^{*})}$, $J_{23}=\frac{\partial dy}{\partial y dt}|_{(x^{*},y^{*},z^{*})}$, $J_{23}=\frac{\partial dy}{\partial z dt}|_{(x^{*},y^{*},z^{*})}$, $J_{31}=\frac{\partial dz}{\partial x dt}|_{(x^{*},y^{*},z^{*})}$, $J_{32}=\frac{\partial dz}{\partial y dt}|_{(x^{*},y^{*},z^{*})}$, $J_{33}=\frac{\partial dz}{\partial z dt}|_{(x^{*},y^{*},z^{*})}$
\newline
\begin{itemize}
    \item  The jacobian at the trivial equilibrium point (0,0,0) is:
    \begin{equation*}
    J(E_{0})=
    \begin{bmatrix}
    \frac{1}{1+c} & 0 & 0\\
    0 & r & 0\\
    0 & 0 & -d
    \end{bmatrix} 
    \end{equation*}
    and the eigenvalues of the Jacobian matrix are
    $\frac{1}{1+c}$, r, $-d$. Since one of the eigenvalues of $J(E_{0})$ is positive, so the system at the equilibrium point $E_{0}(0, 0, 0)$ is always unstable.
    \item The jacobian at $E_{1}$ is: 
    \begin{equation*}
    J(E_{1})=
    \begin{bmatrix}
    \frac{-1}{1+c} & 0 & 0\\
    0 & r & 0\\
    0 & 0 & emax(0,1-\frac{\lambda lqc}{\eta+q})\frac{ql}{\eta+q}-d
    \end{bmatrix} 
    \end{equation*}
    Since r is a positive eigenvalue of the jacobian matrix $J(E_{1})$, $E_{1}$ is unstable.
    
    \item The jacobian at the equilibrium $E_{2}$ is:
    \begin{equation*}
    J(E_{2})=
    \begin{bmatrix}
    \frac{1}{1+c} & 0 & 0\\
    0 & -r & 0\\
    0 & 0 & \frac{(1-q)lK}{\eta+(1-q)K}-d
    \end{bmatrix} 
    \end{equation*}
    The equilibrium $E_{2}$ is unstable as $\frac{1}{1+c}$ is the positive real root of the jacobian.
    
    \item The jacobian at the equilibrium $E_{3}$ is:
    \begin{equation*}
    J(E_{3})=
    \begin{bmatrix}
    \frac{-1}{1+c} & 0 & \frac{lq}{\eta+q+(1-q)K}\\
    0 & -r & 0\\
    0 & 0 & emax(0,1-\frac{ \lambda lqc}{\eta+q+(1-q)K})\frac{ql+(1-q)lK }{\eta+q+(1-q)K}-d
    \end{bmatrix} 
    \end{equation*}
    As the lower diagonal of the jacobian matrix is zero then the roots are $\frac{-1}{1+c}$, -r, $emax(0,1-\frac{\lambda lqc}{\eta+q+(1-q)K})\frac{ql+(1-q)lK}{\eta+q+(1-q)K}-d$. The first two roots are negative so the stability require that third root to be negative. For $\frac{\lambda lqc}{\eta+q+(1-q)K}\geq 1$ the last root is also negative and the system is stable. If $\frac{\lambda lqc}{\eta+q+(1-q)K}< 1$ then the stability require $e(1-\frac{\lambda lqc}{\eta+q+(1-q)K})\frac{ql+(1-q)lK}{\eta+q+(1-q)K}<d$.
    
    \item The uncontaminated prey extinction equilibrium $E_{4}=(x_{4},0,z_{4})$ can be expressed as
    \begin{equation*}
    z_{4}=\frac{(1-x_{4})(\eta+qx_{4})}{lq(1+c)}
    \end{equation*}
    Where $x_{4}$ is the positive root of the quadratic equation
    
    \begin{equation*}
    x_{4}^{2}(eq-e\lambda qc l^{2}-q^{2} d)+{x_{4}}(e\eta ql-2q \eta d)-d {\eta}^{2}=0
    \end{equation*}
    
    The equation has atleast one positive real root if:
    \begin{equation*}
    (e\eta q l-2q\eta d)^{2}+4d(eq-eqcl^{2}-q^{2}d)\geq 0 
    \end{equation*}
    
    \begin{equation*}
    (e\eta ql-2q\eta d) <0
    \end{equation*}
    The jacobian at$E_{4}$ is,
    \begin{equation*}
    J(E_{4})=
    \begin{bmatrix}
    P & Q & R\\
    0 & S & 0\\
    T& U & 0
    \end{bmatrix} 
    \end{equation*}
    The characteristic equation is 
    \begin{equation*}
    (S-E)(E^{2}-PE-RT)=0
    \end{equation*}
    Where,
    \begin{equation*}
    P=\frac{1}{1+c}-\frac{2x_{4}}{1+C}-\frac{\eta lqz_{4}}{\eta+qx_{4}}
    \end{equation*}
    \begin{equation*}
    Q=\frac{lq(1-q)x_{4}}{\eta+qx_{4}}
    \end{equation*}
    \begin{equation*}
    R=-\frac{lqx_{4}}{\eta+qx_{4}}
    \end{equation*}
    \begin{equation*}
    S=-\frac{\eta l(1-q)z_{4}+lq(1-q)x_{4}z_{4}}{(\eta+qx_{4})^{2}}
    \end{equation*}
    \begin{equation*}
    T=\frac{\eta q z_{4}}{\eta+qx_{4}}-\frac{\eta q x_{4}z_{4}+\lambda \eta q^{2} l c x_{4} z_{4}}{(\eta+qx_{4})^{3}}
    \end{equation*}
    $E_{4}$ will be stable if $P<0$ and $T>0$.
    
    \item The contaminated prey extinction equilibrium $E_{5}=(0,y_{5},z_{5})$ can be write as:
    \begin{equation*}
    z_{5}=\frac{r(K-y)(\eta+(1-q)y_{5})}{Kl(1-q)}
    \end{equation*}
    Where $y_{5}=\frac{d\eta}{(1-q)(e-d)}$, the feasibility of the equilibrium require $e>d$.
    The jacobian at $E_{5}$ is
    \begin{equation*}
    J(E_{5})=
    \begin{bmatrix}
    P & 0 & 0\\
    Q & R & S\\
    T& U & 0
    \end{bmatrix} 
    \end{equation*}
    The characteristic equation is 
    \begin{equation*}
    (P-E)(E^{2}-RE-SU)=0
    \end{equation*}
    Where 
    \begin{equation*}
    P=\frac{1}{1+c}-\frac{\eta lqz_{5}}{\eta+(1-q)y_{5}}
    \end{equation*}
    \begin{equation*}
    R=r-\frac{2y_{5}r}{K}-\frac{\eta l (1-q)z_{5}}{(\eta+(1-q)y_{5})^{2}}
    \end{equation*}
    \begin{equation*}
    S=-\frac{l(1-q)y_{5}}{\eta+(1-q)y_{5}}
    \end{equation*}
    \begin{equation*}
    U=\frac{\eta (1-q)z_{5}}{(\eta +(1-q)y_{5})^{2}}
    \end{equation*}
    Therefore, the  equilibrium will be stable if the coefficients of the characteristic equation
    satisfy $P<0$, $R>0$, $SU<0$
    
    \item The coexisting equilibrium $E_{6}=(x_{6},y_{6},z_{6})$ can be found by solving the non-linear equations
    \begin{equation*}
     \frac{1}{1+c}(1-x_{6})-\frac{lqz_{6}}{\eta+qx_{6}+(1-q)y_{6}}=0
     \end{equation*}
     \begin{equation*}
    r(1-y_{6}/K)-\frac{l(1-q)z_{6}}{\eta+qx_{6}+(1-q)y_{6}}=0
    \end{equation*}
    \begin{equation*}
    el(\ 1-\frac{\lambda ql x_{6}c}{\eta+qx_{6}+(1-q)y_{6}} )\ (\ \frac{qx_{6}+(1-q)y_{6}}{\eta+qx_{6}+(1-q)y_{6}} )\ -d=0 
    \end{equation*}

\end{itemize}
\begin{figure}[H]
\hspace*{0cm} 
     \centering
     \begin{subfigure}[b]{0.5\textwidth}
         \centering
         \includegraphics[width=\textwidth]{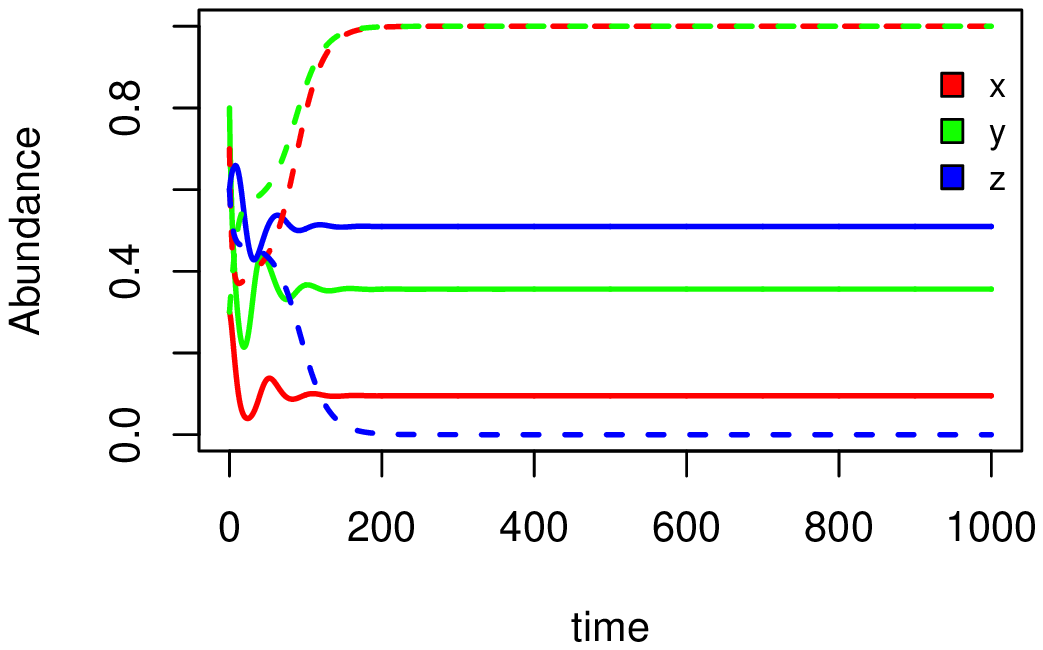}
         \caption{}
         \label{fig:d1}
     \end{subfigure}
     \hspace*{-2cm}
     \hfill
     \begin{subfigure}[b]{0.5\textwidth}
         \centering
         \includegraphics[width=\textwidth]{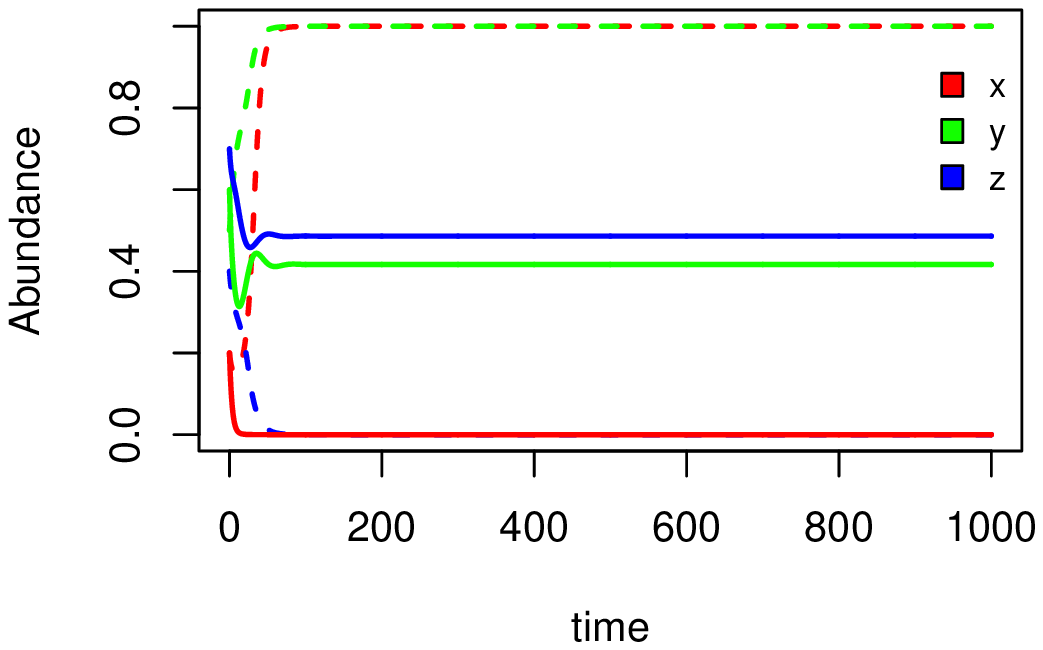}
         \caption{}
         \label{fig:d2}
     \end{subfigure}
     \hfill
     \begin{subfigure}[b]{0.5\textwidth}
         \centering
         \includegraphics[width=\textwidth]{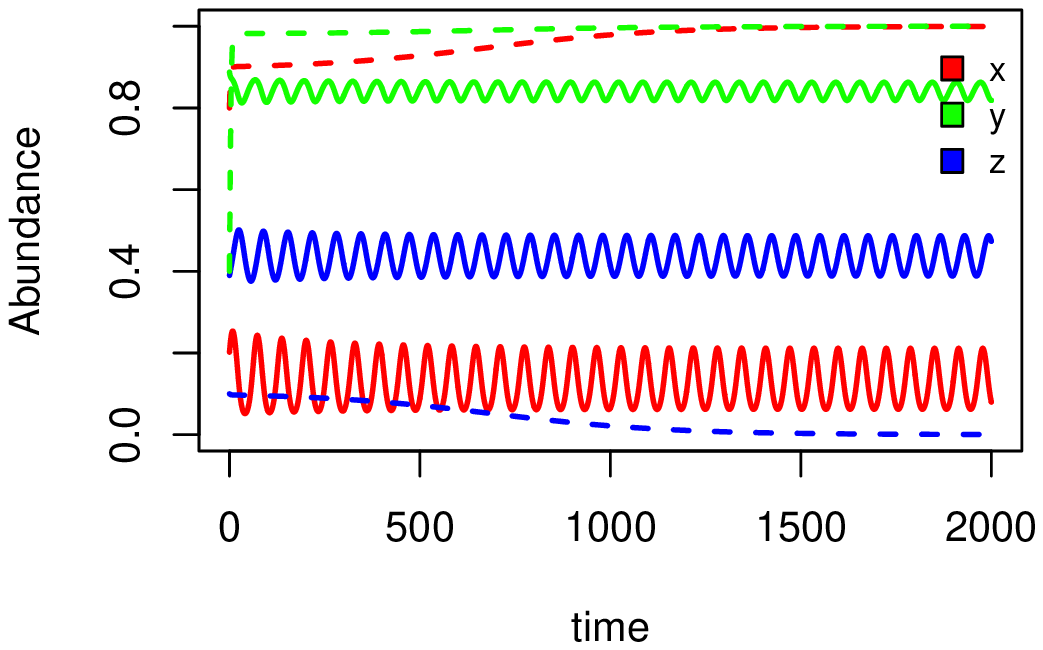}
         \caption{}
         \label{fig:d3}
     \end{subfigure}
     \caption{Time series of abundance at various bistabilisty region where solid and dashed lines denotes two diffrent initial conditions (a) bistability between SC (solid lines) and PDE (dashed lines) in region \textcircled{4},  c=1.3, q=0.55; (b) bistability between PDE (solid lines) and CPE (dashed lines) in region \textcircled{6}, c=0.65, q=0.87; (c) bistability between OC (solid lines) and PDE (dashed lines) in region \textcircled{7}, c=3, q=0.6. Rest of the parameter values are same as in Figure \ref{fig:One}.}
\end{figure}

\end{document}